\documentclass{aa}
\usepackage{natbib}
\usepackage{graphicx}
\usepackage{rotating}
\usepackage{times}
\usepackage{amsmath}
\usepackage{txfonts}
\usepackage{multirow}

\newcommand{\msun}{{\,\rm M}_{\odot}}
\newcommand{\lsun}{{\,\rm L}_{\odot}}

\newcommand{\kms}{\,{\rm km.s}^{-1}}

\newcommand{\nht}{\ifmmode {{\rm NH}_3} \else {NH{\bas 3}} \fi}
\newcommand{\tco}{\ifmmode {^{13}{\rm CO}} \else {$^{13}{\rm CO}$}\fi}
\newcommand{\dco}{\ifmmode {^{12}{\rm CO}} \else {$^{12}{\rm CO}$}\fi}
\newcommand{\cdo}{\ifmmode {{\rm C}^{18}{\rm O}} \else {${\rm C}^{18}{\rm O}$}\fi}

\newcommand{\htco}{\ifmmode {{\rm H}^{13}{\rm CO}^{+} } \else {${\rm H}^{13}
{\rm CO}^{+}$ }\fi}
\newcommand{\hco}{\ifmmode {{\rm H}^{12}{\rm CO}^{+} } \else {${\rm H}^{12}
{\rm CO}^{+}$ }\fi}
\newcommand{\juz}{\ifmmode {{\rm J}=1\rightarrow 0} \else
{J=1$\rightarrow$0}\fi}
\newcommand{\jdu}{\ifmmode {{\rm J}=2\rightarrow 1} \else
{J=2$\rightarrow$1}\fi}
\newcommand{\jtd}{\ifmmode {{\rm J}=3\rightarrow 2} \else
{J=3$\rightarrow$2} \fi}
\newcommand{\jcq}{\ifmmode {{\rm J}=5\!\rightarrow\!4} \else
{${\rm J}=5\!\rightarrow\!4$} \fi}
\newcommand{\as}{\ifmmode {^{\scriptscriptstyle\prime\prime}}
        \else $^{\scriptscriptstyle\prime\prime}$\fi}
\newcommand{\am}{\ifmmode {^{\scriptscriptstyle\prime}}
        \else $^{\scriptscriptstyle\prime}$\fi}
\newcommand{\hh}{\ifmmode {{\rm H}_2} \else {H$_2$} \fi}
\renewcommand{\hco}{\ifmmode {{\rm HCO}^+} \else {HCO$^+$} \fi}
\newcommand{\hhco}{\ifmmode {{\rm H}_2{\rm CO}} \else {H$_2$CO} \fi}
\newcommand{\ddco}{\ifmmode {{\rm D}_2{\rm CO}} \else {D$_2$CO} \fi}
\newcommand{\chhdoh}{\ifmmode {{\rm CH}_2{\rm DOH}^+} \else {CH$_2$DOH} \fi}
\newcommand{\chhhod}{\ifmmode {{\rm CH}_3{\rm OD}^+} \else {CH$_3$OD} \fi}
\newcommand{\chhhoh}{\ifmmode {{\rm CH}_3{\rm OH}^+} \else {CH$_3$OH} \fi}
\newcommand{\tchhhoh}{\ifmmode {^{13}{\rm CH}_3{\rm OH}^+} \else {$^{13}$CH$_3$OH} \fi}
\newcommand{\dcop}{\ifmmode {{\rm DCO}^+} \else {DCO$^+$} \fi}
\newcommand{\cchh}{\ifmmode {{\rm C}_2{\rm H}_2} \else {C$_2$H$_2$} \fi}

\newcommand{\hcccn}{\ifmmode {{\rm HC}_3{\rm N}} \else {HC$_3$N} \fi}

\begin{document}

\title{Sensitive survey for $^{13}$CO, CN, H$_2$CO, and SO in the disks of T Tauri and Herbig Ae stars II: Stars in $\rho$ Oph and upper Scorpius.
\thanks{Based on observations carried out with the IRAM 30-m telescope.
 IRAM is supported by INSU/CNRS (France), MPG (Germany) and IGN (Spain).}
}
\author{
 L.~Reboussin \inst{1,2},
 S.~Guilloteau \inst{1,2},
 M.~Simon \inst{3},
 N.~Grosso \inst{4},
 V.~Wakelam \inst{1,2},
 E.~Di Folco \inst{1,2},
 A.~Dutrey \inst{1,2},
 V.~Pi\'etu \inst{5}}
\institute{
Univ. Bordeaux, LAB, UMR 5804, F-33270, Floirac, France
\and{}
CNRS, LAB, UMR 5804, F-33270 Floirac, France
\and{}
Department of Physics and Astronomy, Stony Brook University, Stony Brook, NY 11794-3800, USA
\and{}
Observatoire Astronomique de Strasbourg, Universit\'e de Strasbourg, CNRS, UMR 7550, 11 rue de l'Universit\'e, 67000 Strasbourg, France
\and{}
IRAM, 300 rue de la piscine, F-38406
Saint Martin d'H\`eres, France
}

\offprints{S.Guilloteau, \email{guilloteau@obs.u-bordeaux1.fr}}

\authorrunning{Reboussin et al.} %
\titlerunning{CN in protoplanetary disks in $\rho$ Ophiuchi}

\abstract
{}
{We attempt to determine the molecular composition of disks around young
low-mass stars in the $\rho$ Oph region and to compare our results with a similar study performed in the Taurus-Auriga region.
}
{We used the IRAM 30 m telescope to perform a sensitive search for CN N=2-1 in 29 T Tauri
stars located in the $\rho$ Oph and upper Scorpius regions. $^{13}$CO J=2-1 is observed simultaneously to
provide an indication of the level of confusion with the surrounding molecular cloud.
The bandpass also contains two transitions of ortho-H$_2$CO, one of SO, and the C$^{17}$O
J=2-1 line, which provides complementary information on the nature of the emission.
}
{Contamination by molecular cloud in $^{13}$CO and even C$^{17}$O is ubiquitous. The CN detection rate
appears to be lower than for the Taurus region, with only four sources being detected
(three are attributable to disks). H$_2$CO emission is found more frequently,
but appears in general to be due to the surrounding cloud.  The weaker emission than in Taurus
may suggest that the average disk size in the $\rho$ Oph region
is smaller than in the Taurus cloud. Chemical modeling shows
that the somewhat higher expected disk temperatures in $\rho$ Oph play
a direct role in decreasing the CN abundance.  Warmer dust temperatures contribute
to convert CN into less volatile forms.
}
{In such a young region, CN is no longer a simple, sensitive tracer of disks, and
observations with other tracers and at high enough resolution with ALMA are required to probe the gas disk
population.
}

\keywords{Stars: circumstellar matter -- planetary systems: protoplanetary disks  -- Radio-lines: stars}

\maketitle

\section{Introduction}

Star formation occurs in clusters that are embedded in collapsing molecular clouds
\citep{Lada&Lada2003}. The different stellar populations and properties, such
as age, mass or density, can have effects on the circumstellar disk evolution,
which strongly depends on the environment. In regions of high stellar density,
disks can be more easily photoevaporated by the intense UV radiation
\citep{Johnstone1998}. This means that the evolution of disks in the vicinity of massive stars will
differ from the evolution of isolated disks \citep[see, e.g.,][]{Mann+etal_2014}.
The disk structure and chemistry
can therefore vary from region to region. 

Circumstellar disks are the sites where planets form and their gas and
dust provide the raw materials for planet building.
Constraining the chemical evolution of protoplanetary disks therefore
is a major
challenge in understanding the planet formation process. Molecular-line
emission is an important tool for deriving the disk characteristics
\citep{Dartois2003, Qi2006}. Because H$_{2}$, the most abundant molecule,
cannot be observed, we rely on lower abundance tracers, principally simple
molecules, to derive those characteristics. In a previous study of $\sim 45$
stars in the Taurus region, we showed that CN N=2-1 is a good tracer of disks
because it is readily detectable in at least 50 \% of the observed sources
\citep[][hereafter Paper I]{Guilloteau+etal_2013}.
Contamination by molecular clouds was minimal in this line; it
was found in fewer than
10 \% of the cases, but some sources exhibited strong lines that most likely
originate from outflows rather than circumstellar disks.

However, the Taurus-Auriga area is a  region of isolated star formation,
and the previous result may not be applicable for other star-forming regions, either
because of stellar density or more simply, because of age. We extend here our study
to the $\rho$ Oph star-forming region, which is both younger and denser than
the Taurus region. We report a search for CN, ortho-H$_{2}$CO, SO, $^{13}$CO,
and C$^{17}$O emissions in 22 young stars located in the $\rho$ Oph region
and 7 stars in the upper Scorpius area.
We attempt to compare these results with previous observations performed
in the Taurus region (Paper I), and more particularly, the CN content in
these star-forming regions.

The paper is organized as follows. The observations and data analysis are
described in Sect. 2. We present the results and the source analysis in
Sect. 3. In Sect. 4 we discuss the difference in molecular composition
observed between the Taurus region
and that of $\rho$ Oph. We conclude in
Sect. 5. An Appendix displays all the observed spectra.

\begin{table*}

\begin{tabular}{lccrrccccccl}
\hline
Object  & $\alpha$ & $\delta$  & $ S_{\nu}(3.4)$ & $ S_{\nu}(1.3)$  & Other & Sp Type & L$_{*}$& Mass & Age & rms & \\
        &  (J2000) & (J2000)  & (mJy)     & (mJy)  & Name  &  &($L_{\odot}$)& ($M_{\odot}$) & (Myr) & (mK)  &\\
\hline
\textit{V1146 Sco} &  15:57:20.0  &  -23:38:50.0 & & & & M0 && & & 28 \\
\textit{J1603-1751} & 16:03:23.7  &  -17:51:42.3 &   && & M2 && & & 36 \\
\textit{AS 205A}  & 16:11:31.4    & -18:38:26.0 & 27.2 & 450   &  HBC 254 & K5 &4.0& & & 67 \\
SR 4     & 16:25:56.1    & -24:20:48.3 & 4.4  &  31  &    AS 206   &   K5  &2.17& 1.14 & 1.1 & 41 \\
GSS 26 ($^{\star}$)   & 16:26:10.3    & -24:20:54.9 & 24.2  &125  &        &M0 &1.39& 0.56 & 0.5 &34 \\
EL 20    & 16:26:18.9    & -24:28:20.2 & 7.3   & 50  &   VSSG 1 &   M0  &0.93& 0.62 & 1.1 & 54 \\
LFAM 1   & 16:26:21.7    & -24:22:50.8 & 17.5  & 250  &  & &  &      &     & 37    \\
DoAr 24E & 16:26:23.4    & -24:21:00.7 & 8.3   &  70  &  GSS 31 & && &    & 35  \\
DoAr 25  & 16:26:23.7    & -24:43:14.1 & 25.0  & 280  & WSB 29 &   K5    &1.43 & 1.12 & 2.1 & 40 \\
EL 24    & 16:26:24.1    & -24:16:14.0 & 48.8  & 390 &  WSB 31 &   K6     &2.58& 0.96 & 0.6 & 38 \\
EL 27    & 16:26:45.0    & -24:23:08.2 & 38.7  & 300 &  GSS 39    &   M0   &0.78  & 0.58 & 1.2 & 38 \\
WL 18    & 16:26:49.0    & -24:38:25.7 & 3.1   &  85  &  GY 129 & K7 &0.3& & &  70 \\
SR 24S   & 16:26:58.5    & -24:45:37.1 & 26.6  & 530  & HBC 262 & K2&4.4& & & 78   \\
SR 21    & 16:27:10.2    & -24:19:12.9 & 4.2   &  95  & EL 30  &  G3    &11.38 & 1.97 & 2.2 & 33 \\
\textit{J1627-2451B} & 16:27:15.1 &  -24:51:38.8 && & & M2 && & & 28 \\
IRS 41   & 16:27:19.3    & -24:28:44.4 & 6.2   & $<60$  & WL 3 &   K7 &1.61    & 0.80 & 0.8 & 50 \\
CRBR 85  & 16:27:24.7    & -24:41:03.2 & 1.5   & 150   &     & &1.4& & & 22    \\
YLW 16c  & 16:27:26.5    & -24:39:23.4 & 6.5   &  60 &  GY 262 &   M1    &1.11 & 0.48 & 0.5 & 23 \\
IRS 49   & 16:27:38.3    & -24:36:58.8 & 4.4   &  25 &  GY 308  &  M0   &1.02  & 0.64 & 1.0 & 47 \\
DoAr 33  & 16:27:39.0    & -23:58:19.1 & 3.7   &  40 &  WSB 53  &   K4    &1.81 & 1.44 & 2.9 & 38 \\
WSB 52   & 16:27:39.5    & -24:39:15.9 & 10.2  &  51 &  GY 314  &   K5     &0.95& 1.04 & 4.2 & 41 \\
IRS 51   & 16:27:39.8    & -24:43:15.0 & 12.7  & 110  &  GY 315 & & & & & 83  \\
{Flying Saucer}($^{\star}$) &  16:28:13.7 & -24:31:39.1 & 3 &  & &  M1 & 0.14 &   &  & 40 \\
WSB 60   & 16:28:16.5    & -24:36:58.0 & 15.3  & 89 &   YLW 58  &  M4  &0.23   & 0.20 & 0.9 & 31  \\
SR 13    & 16:28:45.3    & -24:28:19.2 & 10.0  & 60 &  HBC 266  & & & & & 37 \\
DoAr 44 & 16:31:33.5    & -24:27:37.7 & 10.4  & 105 &   HBC 268 &   K3    &1.55 & 1.29 & 5.1 & 38 \\
\textit{RNO 90}   & 16:34:09.2    & -15:48:16.9 & 7.6   & 25 &   HBC 649 &   G5    &10.24 & 1.87 & 2.3 & 31 \\
\textit{Wa Oph 6} & 16:48:45.6    & -14:16:36.0 & 10.3  & 130   & HBC 653 &   K6  &2.32   & 0.98 & 0.7 & 26 \\
\textit{AS 209} ($^{\star}$)  & 16:49:15.3  & -14:22:08.6 & 17.5 & 300 &  HBC 270&  K5   &2.11  & 1.18 & 1.2 & 24 \\
\textit{HD 163296} ($^{\star}$) & 17:56:21.3  & -21:57:22.0&     &   &          & A1  &  & 2.25 & 5.0 &50 \\
\hline
\end{tabular}
\caption{Source sample. Sources in upper Scorpius are listed
in italics.
Sources marked with ($^{\star}$) are the sources for which CN emission
from disk has been detected. The epoch 2000 coordinates are adopted from SIMBAD.
S$_\nu$ is the continuum flux at 3.4 mm and 1.3 mm. Spectral types and
luminosities (given for $\sim$ 130 pc) are taken from \citet{Ricci2010} except
for AS 205A \citep{Bast2011}, WL 18 \citep{Andrews2010}, SR 24S
\citep{Andrews2010}, CRBR 85 \citep{Pontoppidan2005}, and the Flying Saucer
\citep{Grosso2003}. Masses and ages are derived from the tracks
reported in \citet{Palla&Stahler1999}  using the luminosities and effective
temperatures from \citet{Ricci2010}.}
\label{tab:source}

\end{table*}

\section{Observations and data analysis}
\label{sec:obs}

\begin{table}
\caption{Detected CN emission}
\begin{tabular}{lccc} \hline \hline
Source & Line Flux & Velocity & Width \\
name & (Jy\,km\,s$^{-1}$) & (km\,s$^{-1}$) & (km\,s$^{-1}$) \\\hline
\textit{AS 209}  & 2.1 $\pm$ 0.3 & 4.5 $\pm$ 0.4 & 4.2 $\pm$ 0.7 \\
GSS 26  & 2.2 $\pm$ 0.3 & 4.0 $\pm$ 0.2 & 2.2 $\pm$ 0.4 \\
IRS 41 & 1.5 $\pm$ 0.3 & 5.5 $\pm$ 0.1 & 0.9 $\pm$ 0.2 \\
CRBR 85 & $\approx 2.5$ & \multicolumn{2}{c}{see Fig.\ref{fig:both}} \\
Flying Saucer & 3.4 $\pm$ 0.5 & 4.1 $\pm$ 0.3 & 4.1 $\pm$ 0.6 \\\hline
\end{tabular}
\tablefoot{CN line parameters for the detected sources.}
\label{tab:flux}
\end{table}

\subsection{Source sample}
\label{sec:sub:sample}

Table \ref{tab:source} lists the 30 stars involved in this study. \object{HD 163296},
an isolated HAe star, was used for calibration purposes. Our main
objective was to search for potential targets for
the stellar mass determination, to test models of early stellar evolution. Hence, our
sample is biased toward late spectral types.

\begin{figure}[h]
\includegraphics[width=\columnwidth]{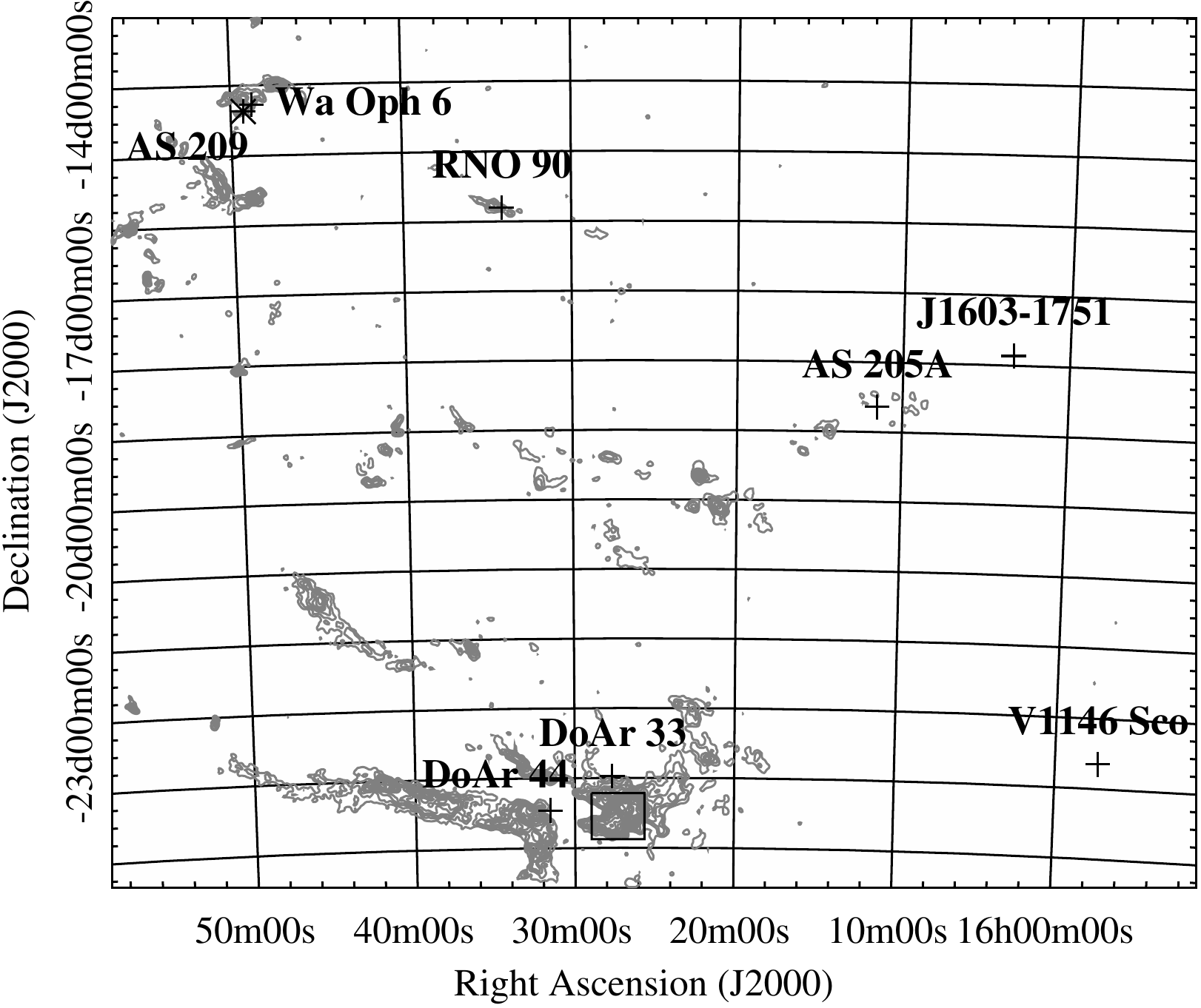}
\includegraphics[scale=0.52]{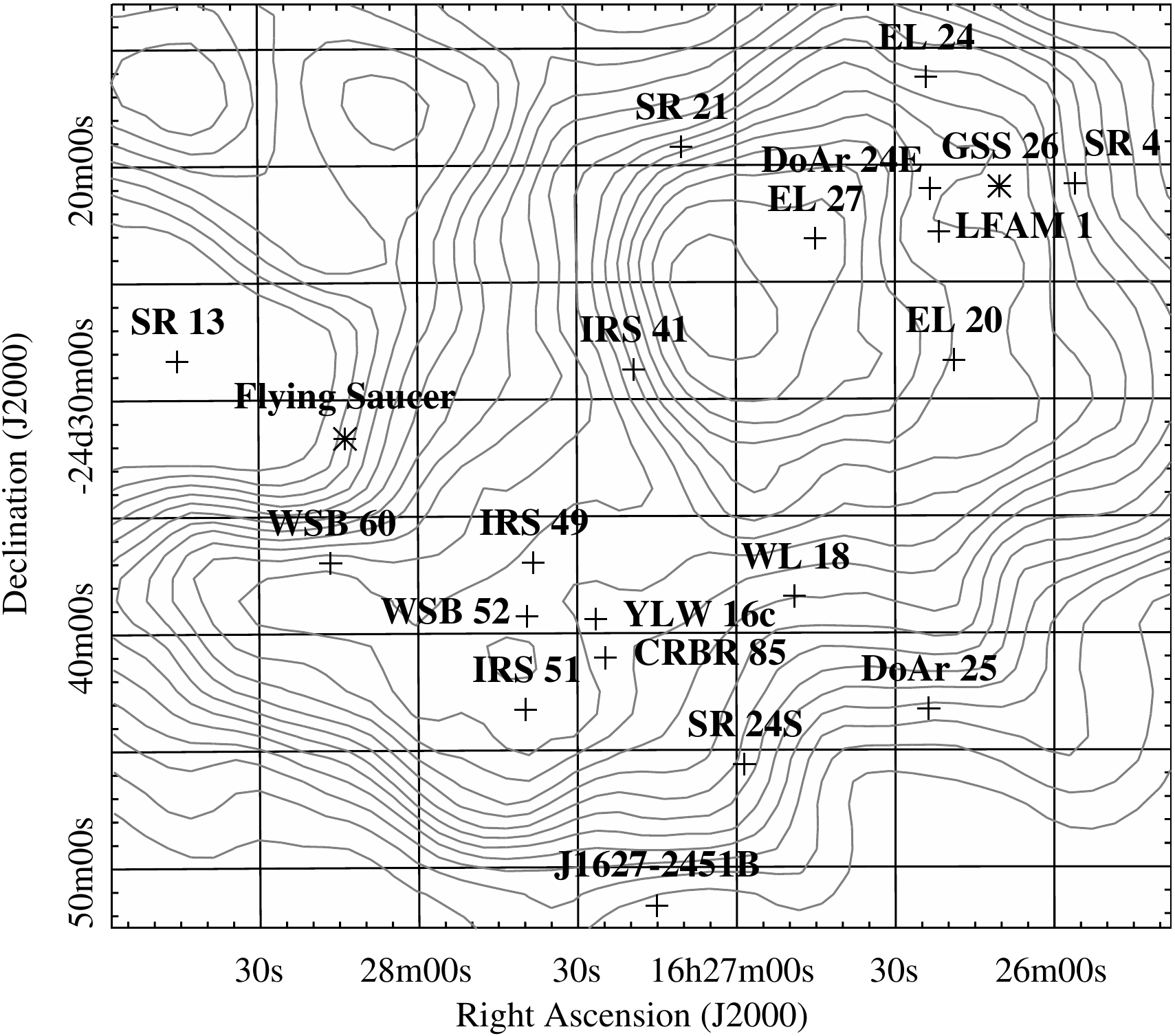}
\caption{Extinction maps of the $\rho$ -Oph star-forming region
\citep[based on the 2MASS PSC][]{Dobashi2013}, contours from 1 to 18 mag
in steps of 1 mag. The bottom map is a zoom on the area in which most sources are located (black square in the top map). Asterisks are
sources with CN detection.}
\label{fig:map}
\end{figure}

The overall sample covers a wide range of spectral types (from M4 to A1), luminosities
(from $0.2$ to $12 \lsun$), and 1.3 mm continuum flux densities
(from $< 4$ to $\sim 800$ mJy), but is not complete with respect to these quantities.
No specific account of extinction and location with respect to molecular clouds
was made when we selected the sources. Sources with outflows were deliberately included.
The distribution of sources is displayed in Fig.\ref{fig:map},
which shows the location of the studied sources overlaid on an extinction
map.  Of these 30 stars, 22 are located in the $\rho$ Oph star-forming region,
for which a reliable distance, 120\,pc \footnote{However, \citet{Loinard2008a}
reported for the two stars WL 5 and VSSG 14 (located near IRS 41) preliminary
measurements indicating a distance of  $\sim 165$\,pc.}, based on VLBI astrometry
\citep{Loinard2008b} or optical extinction \citep{Knude&Hog1998}, is available.
The distances to the other 8 stars might lie in
the range 120 to 170 pc \citep{Loinard2008a, deGeus1989}: we assume
160 pc hereafter.
We report our observations of these stars, but cannot reliably
analyze them in terms of stellar luminosity and mass.

\subsection{Observations}
\label{sec:sub:obs}

Observations were carried out with the IRAM 30 m radiotelescope on October 12-13, 2012,
and from August 8-14, 2013. The weather conditions were quite varied,
with water vapor content changing from 3\,mm on good nights to 6-7 mm on others. Given
the low elevation of the sources (mostly below $25^\circ$), this resulted in
single-sideband system temperatures ranging from 350 K to about 1000 K,
with averaged values between 400 K and 600 K for most sources. However,
pointing conditions were good.

A few strong lines of SO, SO$_2$ , and SO$^+$ found in one of the sources allowed
us to check the sideband rejection, which was found to be around 13 dB.
All observations were made at relatively low but constant elevation. We thus
did not apply any gain-elevation curve correction and used a simple
uniform conversion factor from antenna temperature (T$_A^*$) to flux
density of  $J_K = 9$ Jy/K. A comparison with previous observations on
HD 163296 (Paper I) suggests that our calibration is accurate within 10 \%.

The observations were performed in symmetric
wobbler-switching mode, with the two references $\pm 60''$ away at the
same elevation. The weather conditions were stable enough to provide flat baselines.

As the system temperature varied much from day to day, we observed the
sample of  30 sources (see Table \ref{tab:source}) for
integration times varying between 1 and 8 hours. The resulting sensitivity
at  the nominal resolution of $\delta V \simeq 0.26 \kms$ is given in Table \ref{tab:source} (rms column).
For a typical line width of $\Delta V = 3 \kms$, the corresponding $1 \sigma$ error on
the integrated line flux can be obtained by multiplying this number by
$J_K=\sqrt{\Delta V \delta V} \simeq 8$ Jy.km/s/K.
Our survey depth is thus ranging between 0.18 to 0.65 Jy$\kms$, with a typical
value of 0.3 Jy$\kms$.

\begin{figure*}
  \begin{minipage}[c]{.46\linewidth}
      \includegraphics[width=\columnwidth]{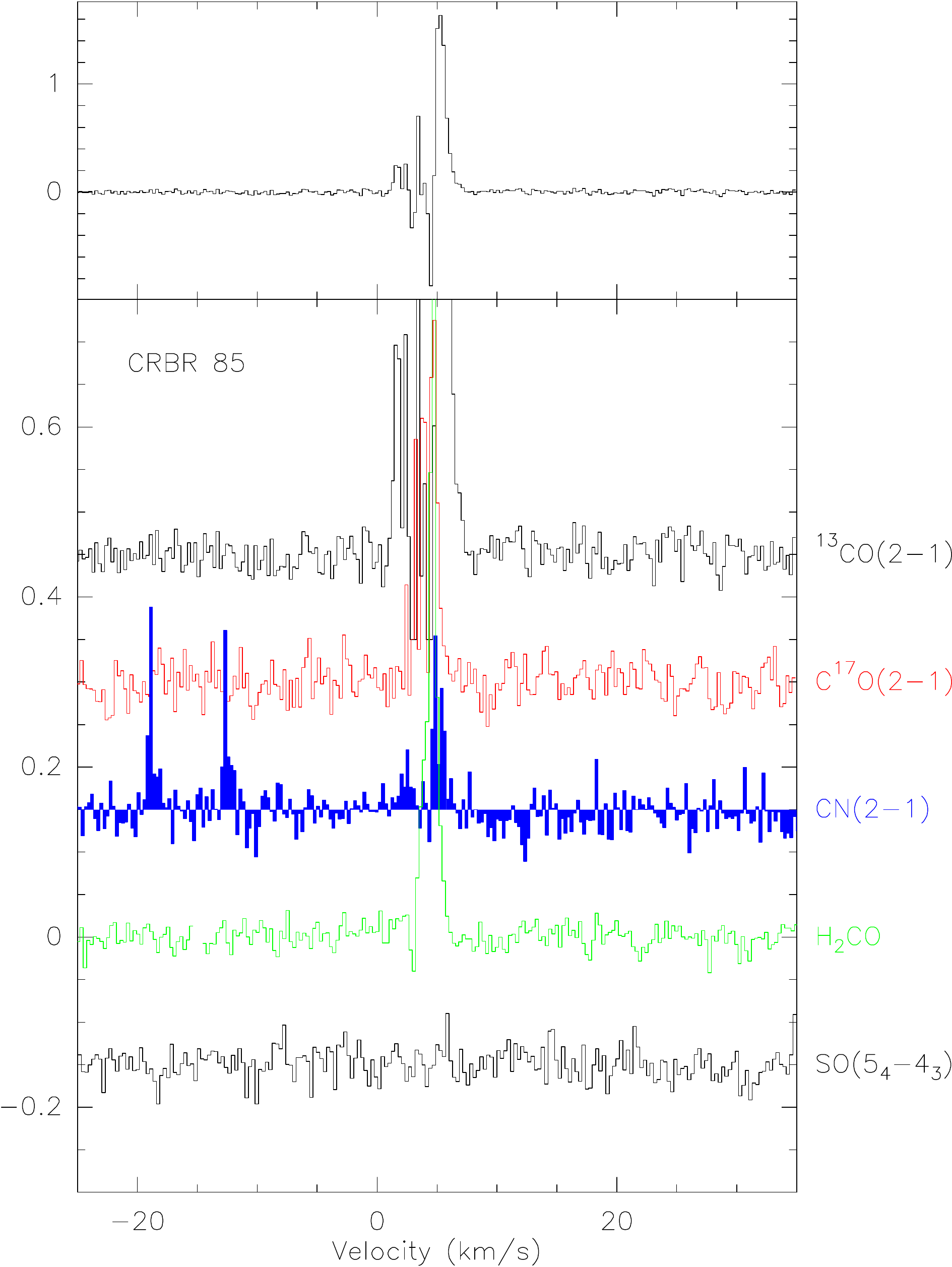}
   \end{minipage} \hfill
   \begin{minipage}[c]{.46\linewidth}
      \includegraphics[width=\columnwidth]{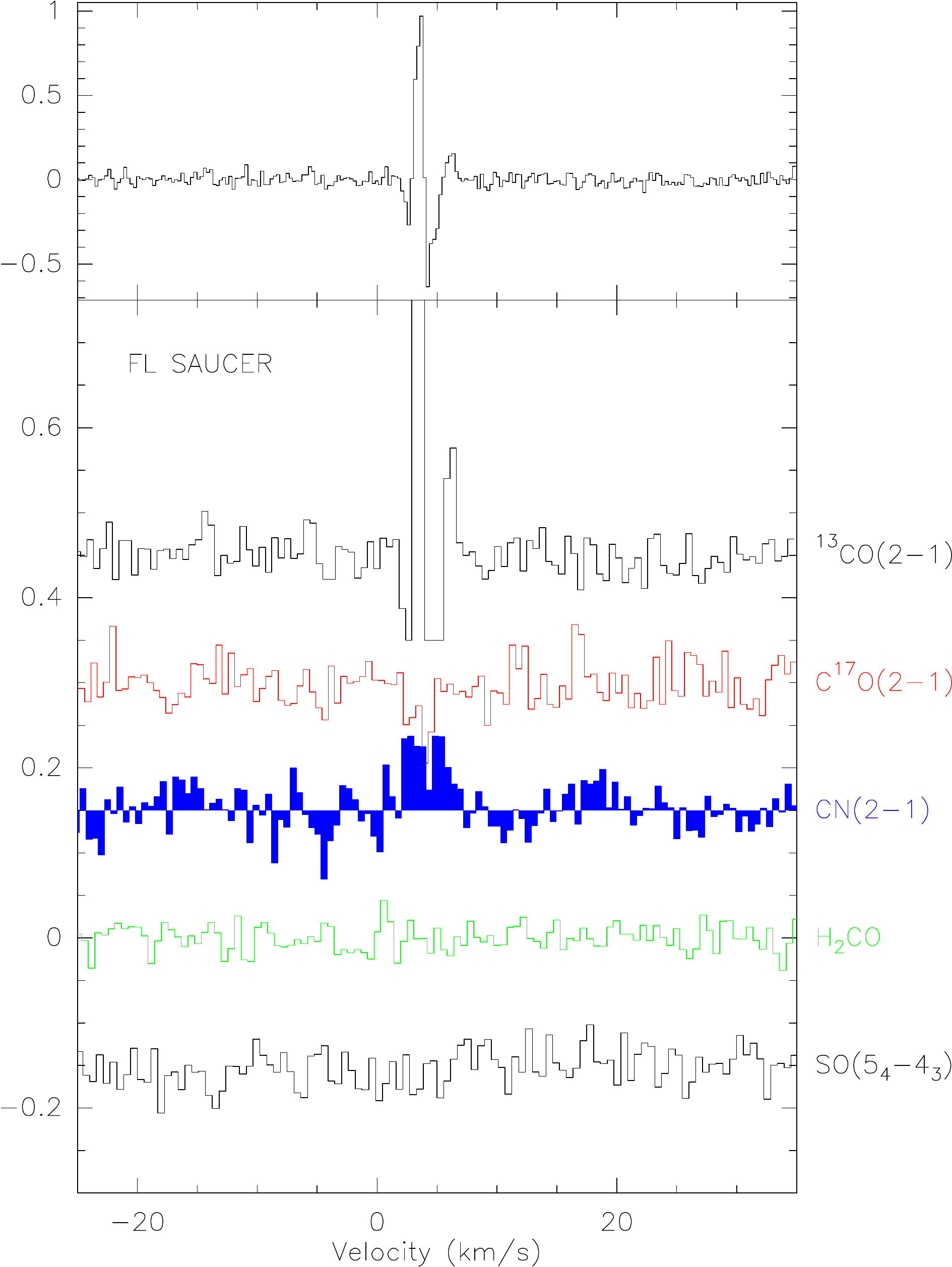}
   \end{minipage}
\caption{Lines toward CRBR 85 and the Flying Saucer.
Top panel: $^{13}$CO J=2-1 spectrum. Bottom panel, from top to bottom: $^{13}$CO J=2-1,
C$^{17}$O J=2-1, CN N=2-1, 
the average of both o-H$_2$CO transitions,
and SO $5_4-4_3$, shifted in intensity by multiples of 0.15 K
to avoid overlap. The intensity scale is the antenna temperature (T$_A^*$ in K):
conversion to flux density can be obtained
using a factor of 9 Jy/K. The spectral resolution is 0.26 km.s$^{-1}$.
For the Flying Saucer, the bottom panel spectra has been smoothed by a
factor 2 for better clarity.}
\label{fig:both}
\end{figure*}

\subsection{Data analysis}
\label{sec:sub:analysis}

We applied the same analysis procedure as in Paper I,
removing linear baselines  in a window $\pm$ 60 km.s$^{-1}$ wide
around each spectral line, and fitting Gaussian profiles to detect lines, with proper account for the hyperfine structure of CN and C$^{17}$O.

The two transitions of ortho-H$_2$CO differ only by the small K splitting,
with the 211.2 GHz line on average $20 \%$ stronger than the 225.2 GHz line.
We computed an average H$_2$CO spectrum and line intensity using
this ratio.


\section{Results}
\label{sec:results}

As described in Paper I, line emission from these objects can
come \textit{\textup{a priori}} from four distinct
regions: molecular cloud(s) along the line of sight, the circumstellar disk,
a molecular outflow, and (especially for the youngest objects) a remnant envelope.
A molecular cloud will exhibit narrow lines, and because of our observing technique,
which differenciates the emission from that $\pm 1'$ away,
may appear in emission or absorption. Disks should lead to symmetric, in general
double-peaked, line profiles, and line widths $\propto \sin{i}$ (with typical
values 2-4 $\kms$, face-on objects being rare).
Contribution from outflows, envelopes, or molecular clouds may vary considerably
and may be difficult to distinguish from disk emission, so that our results
can be dominated by contamination.

Figure \ref{fig:both} shows the spectra toward the two illustrative sources
CRBR\,85  and the Flying Saucer, an edge-on disk at the periphery of $\rho$
Oph \citep{Grosso2003}. Spectra for all other sources
are presented in the Appendix.

The $^{13}$CO results are completely dominated by contamination. Only four sources
show no obvious sign of contamination: AS 205, J1603-1751, DoAr 44, and V1146 Sco.
No disk component can be fit to the observed profiles, except for the
already known case of the isolated HAe star HD 163296.

Even in C$^{17}$O J=2-1, contamination exists, with negative signals
in IRS 41, IRS 49, El 20, SR 21, El 24, El 27, and WSB 60, and strong emission
toward CRBR 85, LFAM 1, YW 16c, and GSS 26. We have a weak detection
in HD 1632926 (see below). 

On the other hand, CN emission is extremely rare. The only emissions that can
be attributed to disks are from HD 163296, AS 209,
the Flying Saucer (see Fig.\ref{fig:both}), and
perhaps GSS 26. Our line flux for AS 209 is slightly lower than that reported
from SMA observations by \citet{Oberg2011}, probably as a result of contamination,
and we also detect the second group of hyperfine components that was below the
sensitivity threshold for the SMA.
We have clear evidence of cloud contamination in DoAR 24E, LFAM 1,
because of negative features, and in CRBR 85.
In the latter case (Fig.\ref{fig:both}), the spectrum is the superposition of
a very narrow component,
with highly anomalous hyperfine ratios, and a broader component that may be
due to an underlying disk.  The case of IRS 41 remains unclear: there is
some CN emission ($1.1 \pm 0.3$ Jy$\kms$), but relatively narrow, $\sim 0.9 \kms$,
and at a velocity where C$^{17}$O shows residual contamination. Line parameters
for the detected sources are indicated in Table  \ref{tab:flux}.

For H$_2$CO, contamination also seems to be rather widespread: negative signals are
observed in El 20, SR 21, SR 4, El 27,  DoAr 24E, and LFAM 1. Positive
signals are detected in GSS 26, WSB 60, El 24, YLW 16c, and CRBR 85. This is a ratio of 11 to 29 sources.

SO is rare: there is strong emission in LFAM 1, and a strong negative signal in
IRS 41, but nothing elsewhere.

\subsection{Specific objects}
\paragraph{\textbf{HD 163296.}}
Although only included here as a calibration source, we note that we
have a $5 \sigma$ detection of C$^{17}$O, with a total line flux
of $4.3 \pm 0.9$ Jy$\kms$. The CN and $^{13}$CO line fluxes
are consistent with previous measurements, indicating good calibration.

\paragraph{\textbf{GSS 26.}}
This is one of the possible new disk detections in CN.  The line
width, however, is very narrow,
$0.75 \pm 0.12 \kms$, and is in fact consistent with that derived from H$_2$CO lines,
$0.63 \pm 0.06 \kms$, although the systemic velocity differs slightly
($3.81 \pm 0.06$ vs $3.43 \pm 0.03 \kms$).
Either this is a disk seen nearly face-on, or the detected lines are
residual contamination by the surroundings, as may be suggested by the
rather strong C$^{17}$O emission.

\paragraph{\textbf{CRBR 85.}}
The spectra of CRBR 85
\citep[also known as CRBR2422.8-3423, see][]{Comeron1993,Pontoppidan2005},
clearly show two components: a very narrow component ($\Delta V \sim 0.5 \kms$)
superimposed on a broader emission ($\Delta V \sim 1.5-2 \kms$). The narrow component exhibits
very unusual hyperfine ratios for CN N=2-1, which, if interpreted as saturation, would indicate
very high opacities. However, they are most likely the result of our differential observation
technique in a cloud extending over 2 arcmin that is nearly opaque in the strongest hyperfine
components, as for CW Tau in Paper I.

The broad component may be attributable to disk or envelope emission. As CRBR 85 is a highly
inclined, embedded object, the limited linewidth indicates a very
low mass star if this emission comes from a disk. The other possibility is emission from a larger distance, that is, from the envelope.
The low line flux would then indicate low excitation, since the detection of weak hyperfine
components indicates substantial opacity. Note, however, that the main group of hyperfine components may be seriously affected
by absorption or contamination due to the narrow component, which
would result in higher
apparent opacities than in reality.

\paragraph{\textbf{LFAM\,1.}}
Also known as GSS 30 IRS 3, this is a strong radio source with emission at 6 cm \citep{Leous1991}. This source is classified as a Class I source \citep{Bontemps2001}.
The spectra in this direction are very unusual. We have strong emission from SO and SO$_2$ , but
also the ion SO$^+$, as well as several isotopologues of SO and SO$_2$. H$_2$CO is quite bright,
and CN shows a negative feature that indicates contamination by a molecular cloud.
S-bearing molecules are in general enhanced in shocks. In fact, SO$^+$ is a clear indicator
of shock chemistry \citep{Turner1992}, predicted to be abundant
in dissociative shocks \citep{Neufeld&Dalgarno1989}, and has recently been detected in
the jet-driven outflow of L1157 \citep{Podio2014}.  The unusual spectrum
of LFAM\,1 therefore probably indicates an outflow
from a very young  object. Another possibility is some contribution from an outflow
driven by GSS 30 IRS 1 because LFAM 1 is located 15$^{\prime\prime}$ to the northeast
of this Class I source, on the edge of its extended V-shaped nebula
\citep[see Fig. 8 of][]{Allen2002}, typical of a large outflow cavity.

\paragraph{\textbf{YLW 16c.}}
This source is the only one clearly showing the double-peaked spectrum
expected from Keplerian rotation in one molecule, here in H$_2$CO. However,
the H$_2$CO/CN ratio is high. In Paper I, we only had strong H$_2$CO
emission and/or high H$_2$CO/CN ratio from sources with outflows, or
relatively massive stars, while YLW 16c is of spectral type M1.

\section{Discussion}
\label{sec:discussion}

\subsection{$\rho$ Ophiuchi and Taurus: two different regions}
The properties of sources in $\rho$ Oph look very different
from those  in Taurus-Auriga: only two potential disk detections in CN (GSS 26,
Flying Saucer), plus perhaps CRBR\,85 and  one source in SO (LFAM 1)
out of 22 sources. In Taurus,
there are 21 disks detected in CN
(this number excludes the sources that also have SO emission of unknown origin)
out of 46 sources, and 8 sources emitting in SO out of 41.

Our survey depth ranges between 0.18 to 0.65 Jy$\kms$, with a median
value of 0.32 Jy$\kms$, a mean of 0.36, and a dispersion of 0.17.
Rescaled to the same distance (i.e., corrected by a factor $(120/140)^2$), the Taurus
observations of Paper I reached a median sensitivity of 0.20 Jy$\kms$ (mean 0.25) with a
dispersion of 0.17. There is thus a typical factor of 1.5 in sensitivity.\\

\begin{figure}[t]
\centering
\includegraphics[width=\columnwidth]{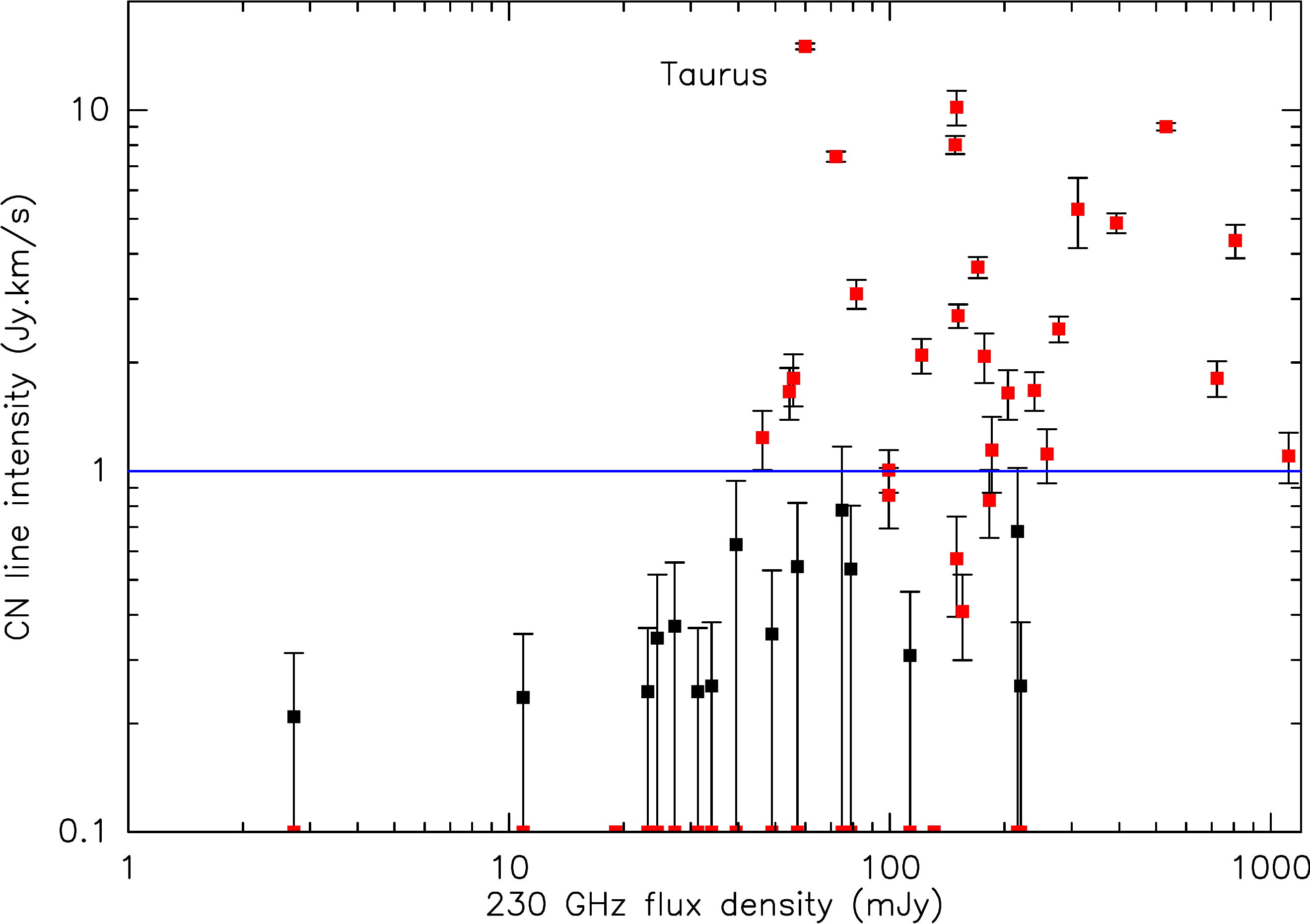}
\includegraphics[width=\columnwidth]{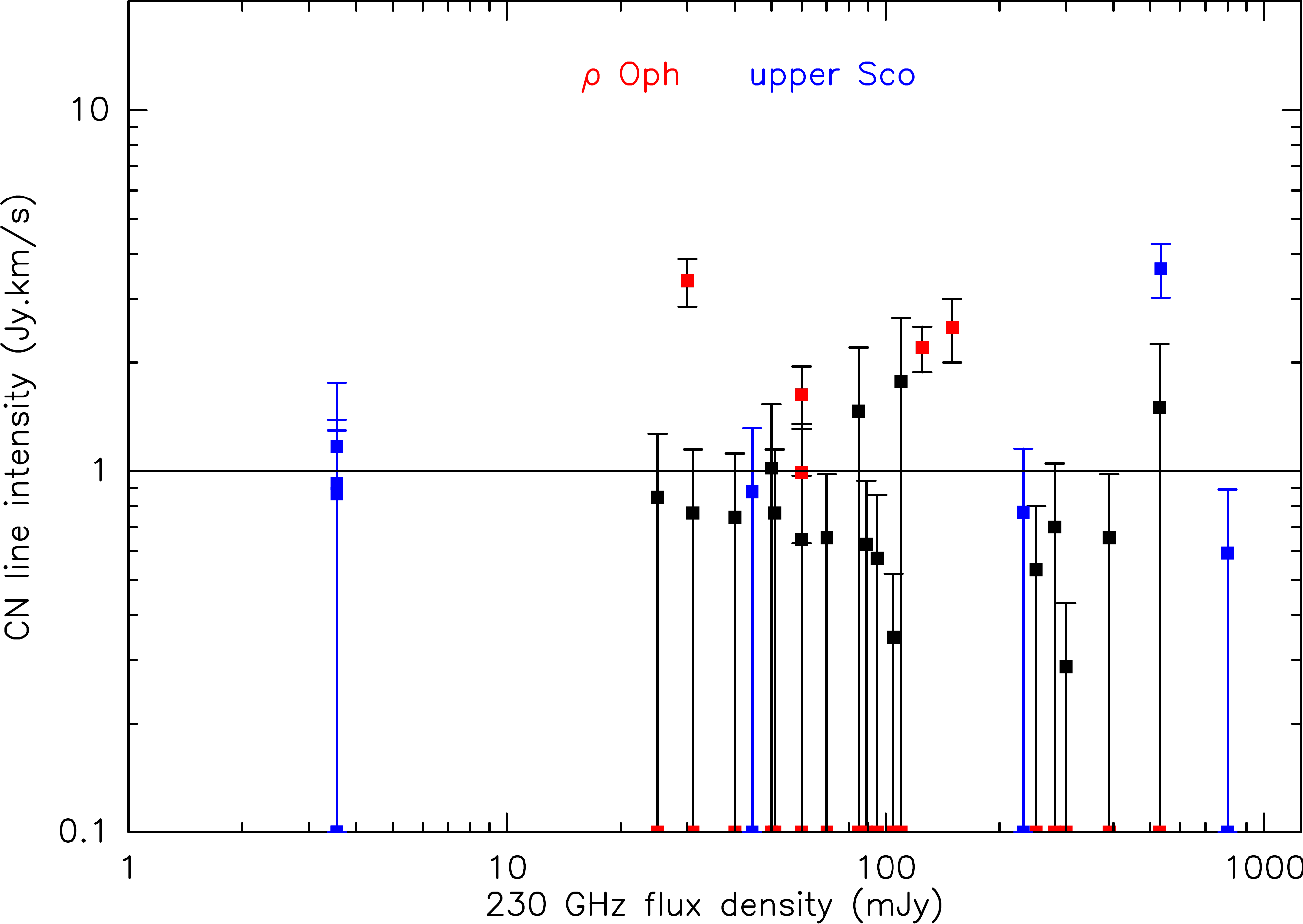}
\caption{CN line intensities vs 1.3\,mm continuum flux. Red dots highlight
detected sources in $\rho$ Oph and Taurus, while blue is for sources in upper
Scorpius. All fluxes have been rescaled to a distance of 120 pc.}
\label{fig:cn230}
\end{figure}

\begin{figure}[ht]
\centering
\includegraphics[width=\columnwidth]{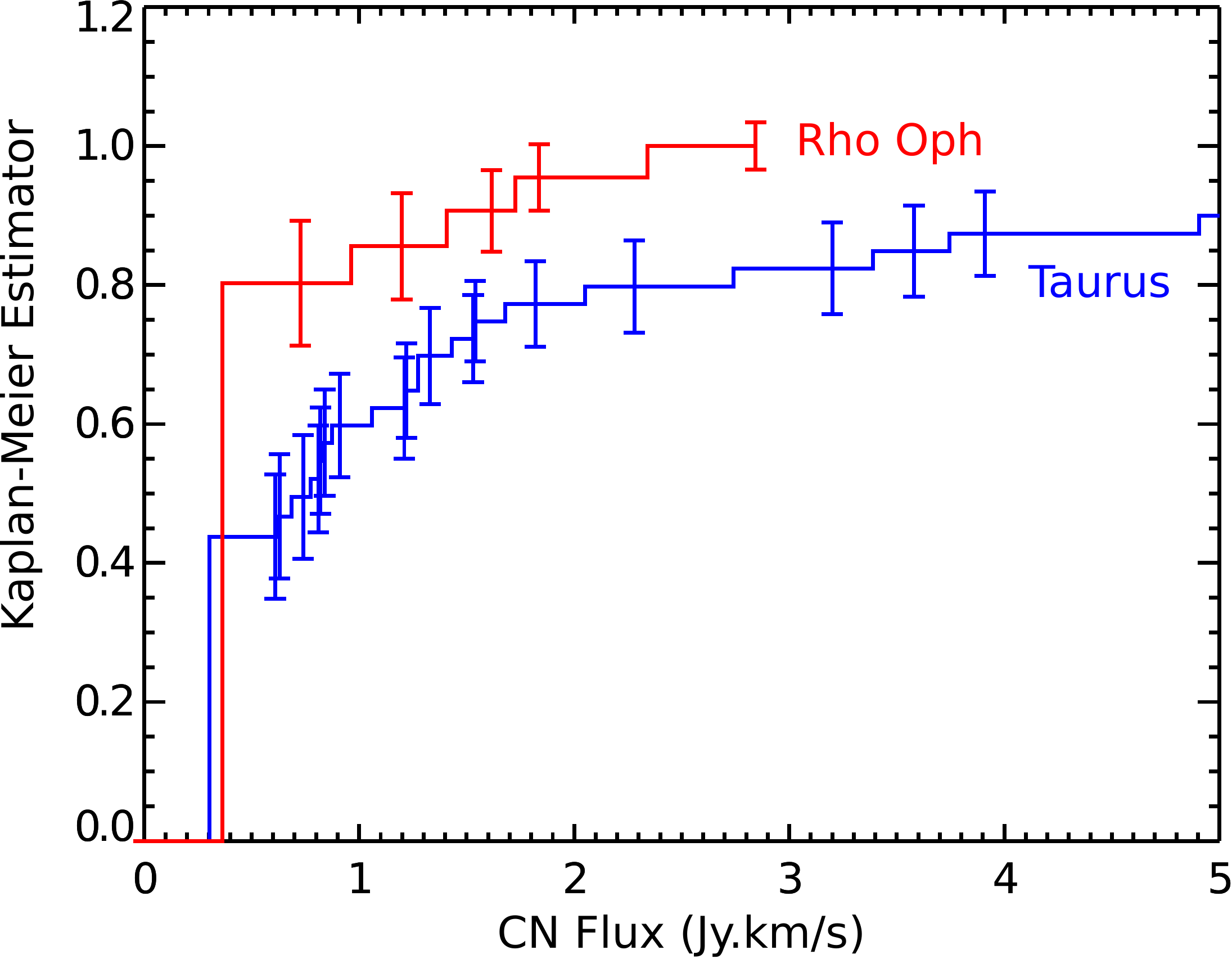}
\caption{Cumulative flux density distribution for CN emission in
Taurus (blue line) and $\rho$ Oph (red line), constructed using the
Kaplan-Meier estimator to include upper limits and scaled to a common distance (d=120 pc).
The sample size are 46 stars for Taurus (Paper I) and 22 stars for $\rho$ Oph.}
\label{fig:kaplan}
\end{figure}

However, this does not appear to be the cause of the large apparent difference in
detection statistics between the two regions. Figure \ref{fig:cn230} shows the CN line
intensity vs.
the 1.3 mm continuum flux (rescaled to a common distance of 120 pc) for both regions.
Red indicates that CN has been detected. The Taurus region exhibited a high disk detection
rate ($>$ 50 \%), whereas only three or four sources have been detected in the $\rho$
Oph region. Furthermore, the population of bright disks found in Taurus has not been
detected in our $\rho$ Oph study.

We also performed a more detailed study of the CN flux density distribution in both regions.
We applied a statistical treatment to our data that takes upper limits into account.
We used the Kaplan-Meier estimator, which provides a maximum likelihood distribution for a randomly censored dataset. A part of our data being limited (upper limits), this estimator allows us to estimate the fraction of sources for which the flux is below (or above) a given value. The details about this
method can be found in \citet{Feigelson&Nelson1985}. This estimator only requires that
the censoring be random, which is the case in our samples since we have considered sources
lying at different distances. We computed the Kaplan-Meier estimator for detections and
nondetections (upper limits) for both samples. The results are displayed in Fig.\ref{fig:kaplan}.
The distributions show the probability that a given source in our sample has a CN flux
higher than the abscissa value. The difference between the two samples is large:
for the Taurus region, the median value for the flux is $\sim$ 0.8 Jy$\kms$, whereas this value is about twice lower for the $\rho$ Oph sample.
Only 20 \% of the sources have a flux higher than 1 Jy$\kms$ in $\rho$ Oph,
whereas the sources in Taurus are much brighter since 40 \% of the sources
have a flux higher than 1 Jy$\kms$.\\

\subsection{Possible cause of these differences}
The CN disk detection rate clearly is much lower in the $\rho$ Oph region than
in the Taurus region. Furthermore, we observed a lack of bright sources in $\rho$ Oph.
We discuss in this section some hypotheses that might explain the differences
between the two regions.

\paragraph{\textbf{Different stellar properties (sample bias)?}}
As discussed in Sect.\ref{sec:sub:sample}, the two samples were not built using
the same criteria. Unlike the Taurus sample, which included a variety
of stars from spectral type M4 to A0, our $\rho$ Oph sample mostly
includes stars in the M2-K3 range and has no warm star (only one G5 and
one G3). However, CN is most easily detected precisely in this M2-K3 range in Taurus sources (Paper I). If we restrict the
Taurus sample to these spectral types, the difference between $\rho$ Oph
and Taurus is even more striking.

\paragraph{\textbf{Contamination?}} Might contamination be the only cause of the difference?
While it is obviously stronger in this region, this looks very unlikely.
Contamination by clouds is best traced by H$_2$CO and C$^{17}$O, which always
show very narrow lines. Such narrow lines would not have prevented a detection of
the wider lines coming from Keplerian disks in CN.

\paragraph{\textbf{Smaller disks?}} In Paper I, we suggested that CN surface
densities were essentially constant (except in the warmer sources). Under this
assumption, which was substantiated by
chemical modelling \citep{Chapillon2012}, the CN line flux provides a nearly
direct measure of the disk size.
The apparent radii derived from the line flux under this simple hypothesis agreed
well with interferometric determination in Taurus. With the same assumptions,
the measured flux densities indicate here that
we have detected three disks with outer radii 500 - 600 au (GSS 26, AS 209, and the Flying Saucer),
have $3 \sigma$ upper limits around 500 au for three other sources (WL 18, SR 24S and IRS 51),
and that most sources are smaller than 320 au, with only one putative disk at 450 au,
IRS 41.

The smaller size observed here may be due to the source ages,
disks in $\rho$ Oph being younger and therefore having less
time to spread by viscous effects.
Making this statement quantitative is difficult, however, because
the age difference between $\rho$ Oph \citep[age $< 1$ Myr][]{Luhman+Rieke_1999}
and Taurus \citep[$\sim 1$ Myr, but with sources up to a few Myr][]{Luhman+etal_2003}
is not well known. In the framework of self-similar viscously spreading disks \citep{Lynden-Bell+Pringle_1974},
the disk radius evolves with time as $T^{1/(2-\gamma)}$,  where $\gamma$ is the
radial exponent of the viscosity and $T = 1+t_*/t_s$, $t_*$ being the disk age
and $t_s$ the viscous timescale.  The classical $\alpha$ disk prescription
yields $\gamma = 3/2-q$ , where $q$ is the radial exponent of the temperature
law, thus $\gamma \sim 1$ as $q \sim 0.5$ for typical disks
\citep[see, e.g.,][for derivations]{Guilloteau+etal_2011}. Viscous timescales are
unfortunately poorly known. Given all uncertainties, it is possible that
viscous spreading in disks located in the Taurus region produces disk sizes larger by 1.2 to 2 times than those in $\rho$ Oph. Even a limited size difference might be sufficient
to explain the observed distribution: according to Fig.\ref{fig:kaplan}, 80 \% of the sample
sources have a flux below 1 Jy$\kms$ in $\rho$ Oph, while for Taurus this
is obtained for 2 -- 3 Jy$\kms$, which means that radii differing by 1.4 -- 1.7 could bring the
two distributions to agree reasonably well. $\rho$ Oph would still lack the
brighter sources found in Taurus, but these are presumably outflows, not disks.

The smaller size might also be a result of initial conditions linked to the
more crowded environment in $\rho$ Oph, an effect that is known
to affect the size of the protostellar envelopes \citep{Motte1998}. The missing
protostellar material at large distances is the one with the highest angular
momentum, which means that disks may be born smaller in $\rho$ Oph.

\begin{figure*}
\begin{minipage}[c]{.5\linewidth}
\includegraphics[width=\columnwidth]{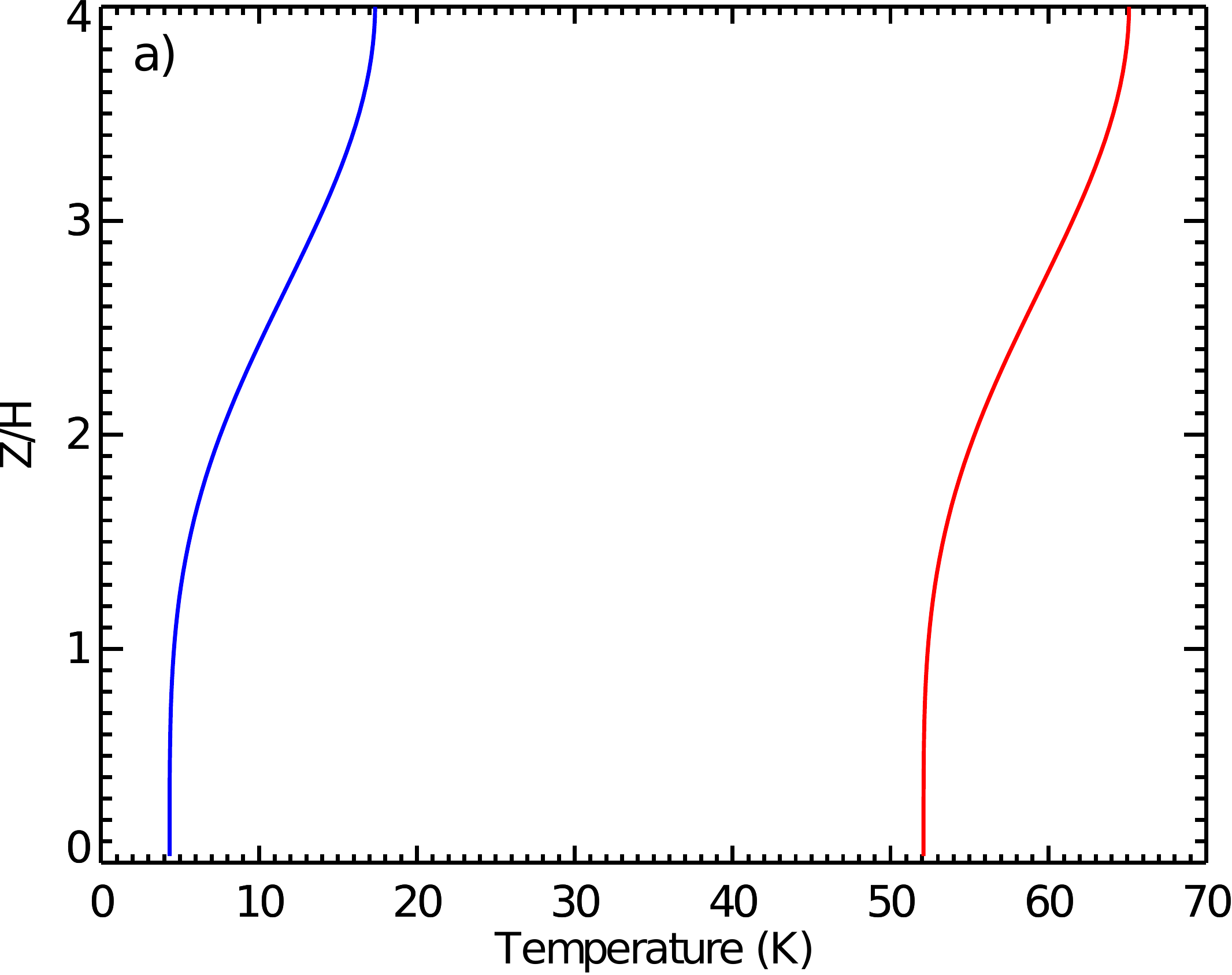}
\end{minipage}
\hfill
\begin{minipage}[c]{.5\linewidth}
\includegraphics[width=\columnwidth]{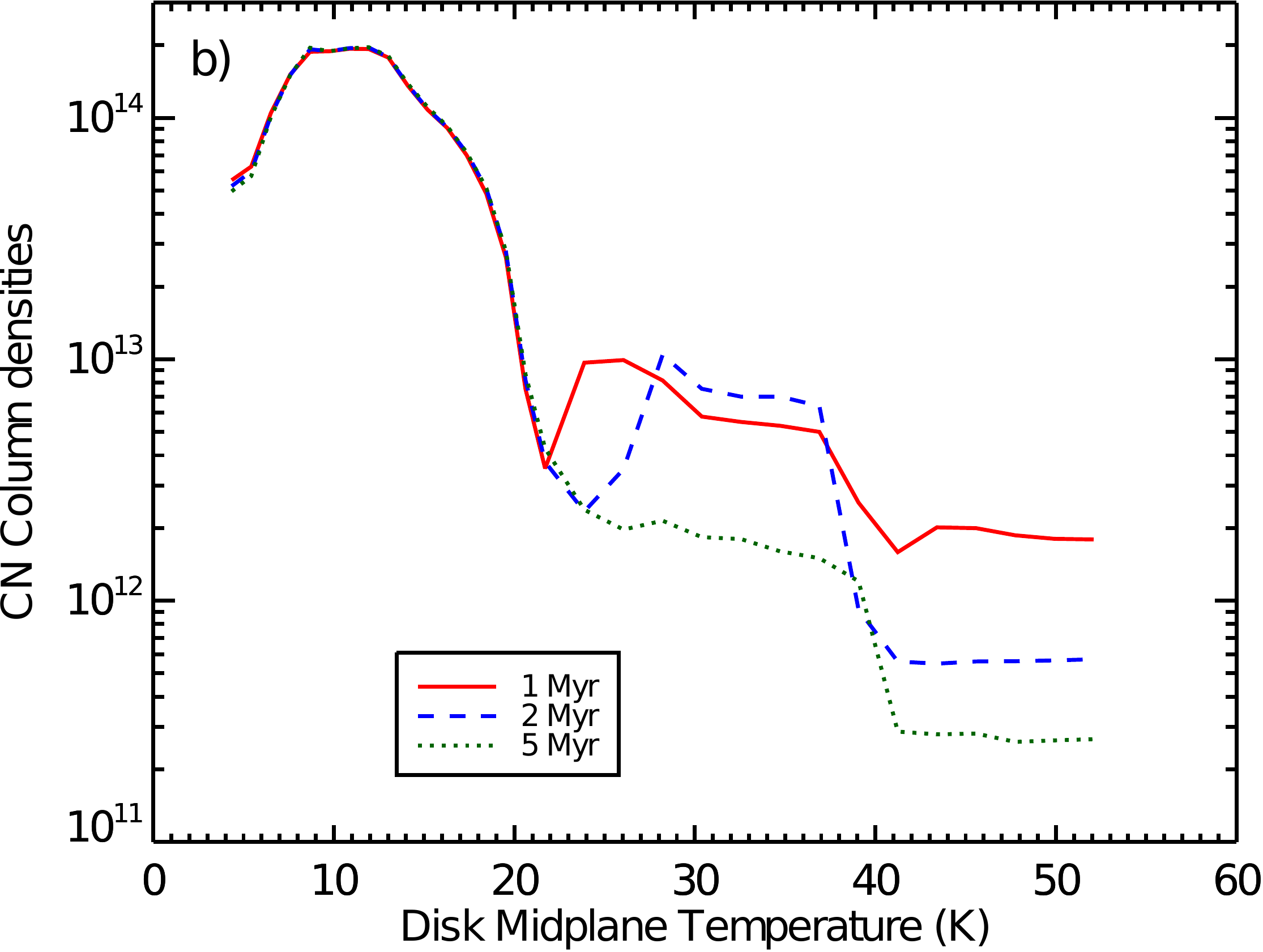}
\end{minipage}
\hfill
\begin{minipage}[c]{.5\linewidth}
\includegraphics[width=\columnwidth]{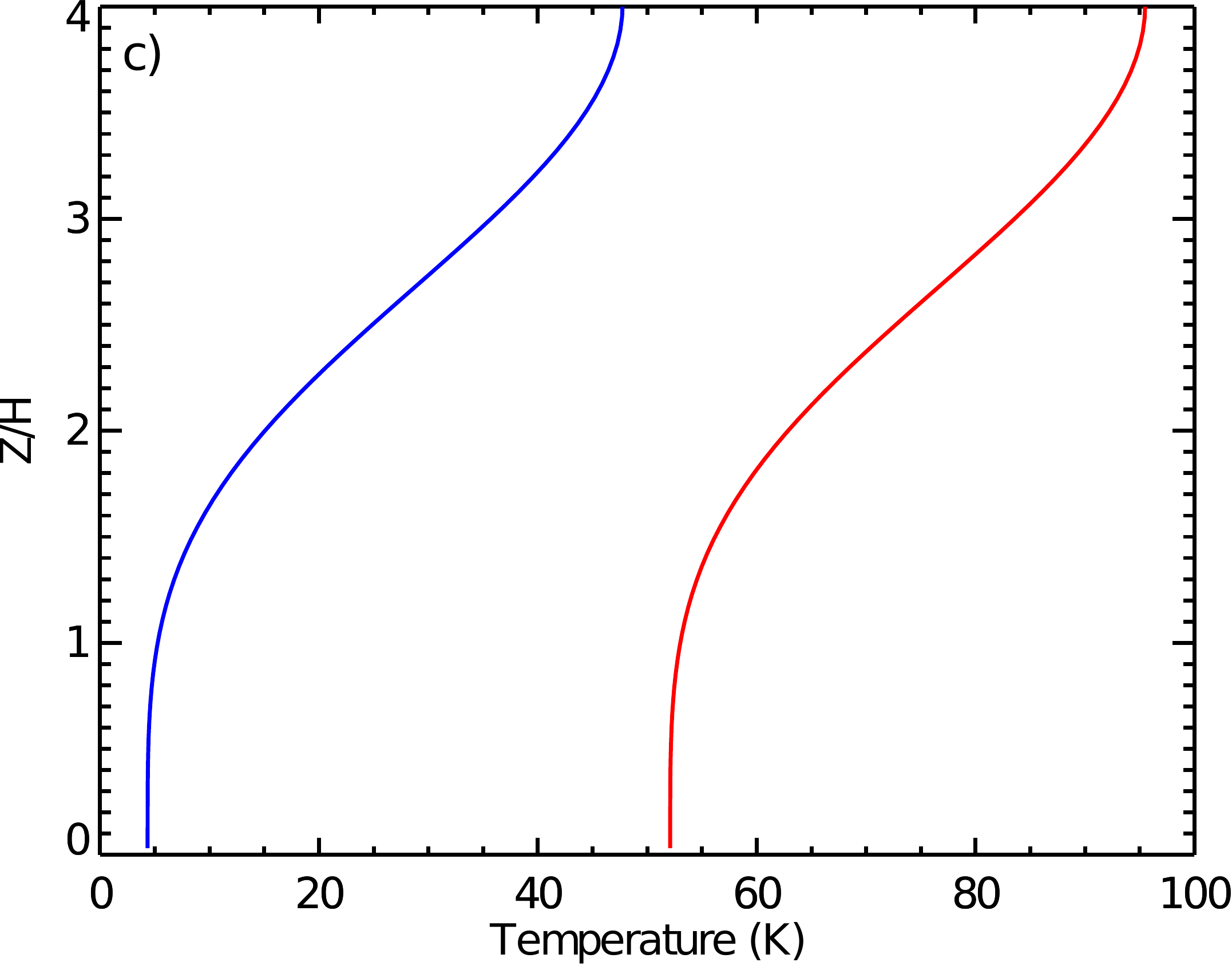}
\end{minipage}
\hfill
\begin{minipage}[c]{.5\linewidth}
\includegraphics[width=\columnwidth]{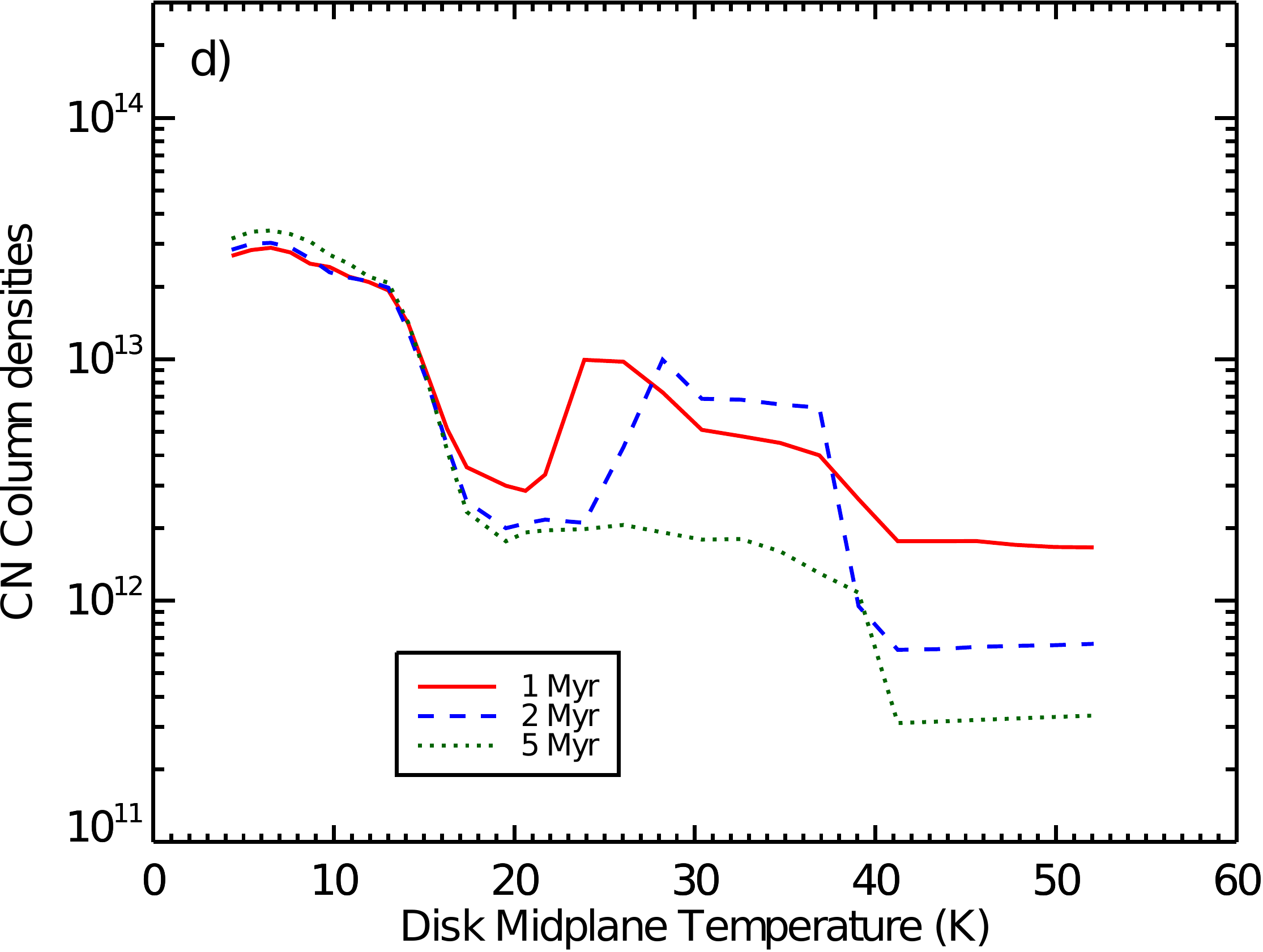}
\end{minipage}
\caption{Panel (a) represents the two extreme
temperature profiles at 300 au for a weak temperature gradient, and
the corresponding CN column density as a function of the
disk midplane temperature for three different ages: 1, 2, and 5 Myr
are shown in panel (b). Panel (c,d): as (a,b), but
for a strong temperature gradient.}
\label{fig:cn}
\end{figure*}

\paragraph{\textbf{Different CN chemistry?}}
The intrinsic differences in disk properties between $\rho$ Oph and Taurus
can affect the CN chemistry and invalidate the assumption of similar
CN surface densities, however. Because  $\rho$ Oph is younger, the stars are brighter: from
evolutionary tracks, a factor of about 2-3 in luminosities for
stars between 1 and 3 Myr is expected. This in turn should lead to a factor $(2-3)^{1/4} = 1.2-1.3$
in disk temperature at the same radius. Moreover, disks are not at chemical equilibrium,
therefore age matters. In the next section, we present the results we obtained
with our chemical model that evaluates the importance
of these effects.

\subsection{Impact of the chemistry}
We used the Nautilus gas-grain model described
in \citet{Semenov2010} and \citet{Reboussin2014}. Nautilus computes the abundance of
species as a function of time in the gas phase and at the surface of the grains. The
chemical network contains 8624 reactions: 6844 are pure gas-phase reactions 1780 are
grain-surface and gas-grain interactions. The gas-phase network used for this work is the same as described in \citet{Reboussin2014}. The full network is available on the KIDA (KInetic
Database for Astrochemistry) website \footnote{http://kida.obs.u-bordeaux1.fr/models}. The dust grains are represented by spherical
particles with a radius of 0.1 $\mu$m and are made of amorphous olivine. The gas and
dust temperatures were assumed to be the same, and we used a dust-to-gas mass ratio of 0.01.
To obtain the initial chemical composition of the disk, we first computed the chemical
composition of the parent molecular cloud. We ran Nautilus during 10$^{6}$ yr for
typical dense cloud conditions: a gas density of 2$\times$10$^{4}$ cm$^{-3}$, a temperature
of 10 K, a visual extinction of 10, a cosmic ray ionization rate of 1.3$\times$10$^{-17}$ s$^{-1}$, and a C/O ratio of 0.5. The elemental abundances used for this work are listed in
\citet{Reboussin2014} (see Table 1 of their paper), except for the oxygen elemental
abundance, for which we considered the value 3.3$\times$10$^{-4}$ (a low-depletion case).
For the disk physical parameters, we used the parametric disk
model described in \citet{Hersant2009} except for the vertical temperature profile, which
is based on \citet{Williams&Best2014}. The model corresponds to a disk mass
of 0.03 $\msun$ with a 1/$r^{1.5}$ surface density profile out to $r=700$ au.
The disk mid-plane temperature was used as a free parameter.
The abundances of chemical species in the disk  were calculated up to 5$\times$10$^{6}$ yr.

Figure \ref{fig:cn} shows the CN column densities as a function of the disk midplane
temperature at a radius of 300 au, at three different disk ages and for two different
temperature profiles (small or strong vertical gradient).
Overall, the CN column density
decreases as the temperature increases and even reaches values well below any detectable
levels (down to a few 10$^{11}$ cm$^{-2}$) at temperatures above 40 K
in the disk midplane. With the 30 m radiotelescope, our detection limit is
around 3$\times$10$^{12}$ cm$^{-2}$ for a disk of 400 au radius
for a detection level of 1 Jy km.s$^{-1}$. This value does not vary
much with the temperature.
The temperature dependence is complex and non-monotonic because
the dominant reaction routes for CN change with temperature.
The gas-phase CN abundance is mostly affected by grain-surface
chemistry: at higher temperatures, the diffusion of the species at the surface of
the grain is much more efficient, which contributes to convert molecules on
grains into less volatile
forms. As this conversion process is relatively slow, the temperature
dependence is more pronounced at later ages. The difference between the two
temperature profiles is quite small at high temperatures, but for very
low temperatures (below 15 K), the CN column density is lower
when we use a strong temperature gradient.
These results agree with the observations obtained in Paper I, in which
CN was undetectable in most warm sources.
As mentioned before, all these simulations were performed considering
the same temperature for the gas and dust.
Models in which these two quantities differ indicate that
the decisive factor for CN chemistry is the dust temperature.
Figure \ref{fig:cn} also shows that the 20-30 \% higher temperature of disks in $\rho$ Oph
is sufficient to affect the CN surface density by a factor of a few, except for very
low temperatures (below 10 K). Similar differences exist as a function of disk ages for
disk temperatures above 15 K.\\


\section{Conclusions}
\label{sec:summary}

We have performed a sensitive survey of 29 young stars in the $\rho$ Oph
and upper Scorpius
regions in CN, ortho-H$_2$CO, SO, $^{13}$CO, and C$^{17}$O rotational lines near
206 - 228 GHz with the IRAM 30 m telescope. Compared to a similar study performed
for the isolated
star formation region Taurus-Auriga, the detection rate of CN is much lower in
the $\rho$ Oph region.

This result may indicate that disks in the $\rho$ Oph region are on average smaller than
those in the Taurus-Auriga complex, perhaps because they have not spread out sufficiently
by viscous diffusion, or as a result of initial truncation of their parental protostellar
condensation.

However, the CN chemistry is shown to be sensitive to disk temperature, so that a direct
comparison of the disk properties would require resolving the disks out instead of relying
on the good correlation between CN flux and disk radii previously found in the Taurus region.
Interferometric observations are required for this purpose.

Finally, these results show that even large single-dish telescopes are severely limited in
identifying disks around embedded young stars because of contamination, but also because
of sensitivity. Disks smaller than about 300 au, which represent 50 \% of the disks
in the sample studied by \citet{Guilloteau+etal_2013} in the Taurus region, are beyond
the sensitivity limit of even the largest telescope operating at 1.3 mm, the IRAM 30 m. Only
ALMA is sensitive enough in the southern hemisphere to detect a substantial fraction
of the gas disk population.

\begin{acknowledgements}
We acknowledge all the 30 m IRAM staff for their help during the observations.
This work was supported by ``Programme National de Physique Stellaire'' (PNPS) and ``Programme
National de Physique Chimie du Milieu Interstellaire'' (PCMI) from INSU/CNRS. The work of MS
was supported in part by NSF grant AST 09-07745.
This research has made use of the SIMBAD database,
operated at CDS, Strasbourg, France.
VW's research work is funded by the ERC Starting Grant (3DICE, grant agreement 336474).
\end{acknowledgements}

\clearpage

\bibliography{biblio}
\bibliographystyle{aa}
\clearpage
\appendix

\section{Spectra for individual sources}

This appendix displays the spectra toward the various sources (in the
same order as listed in Table \ref{tab:source}). For each
source, the top panel shows the $^{13}$CO J=2-1 spectrum.
The bottom panel displays on a common  scale from
top to bottom the spectra of $^{13}$CO J=2-1 (with fit as in top panel),
C$^{17}$O J=2-1, CN N=2-1, 
the average of both o-H$_2$CO transitions,
and SO $5_4-4_3$, shifted in intensity by multiples of 0.15 K
to avoid overlap. The intensity scale is the antenna temperature (T$_A^*$ in K):
a conversion to flux density can be obtained
using a factor of 9 Jy/K.

The spectral resolution is $0.26 \kms$ in the upper panel, while the other
spectra have been smoothed by a factor 1, 2, or 4 for better clarity.

\label{app:spectra}

\clearpage
\begin{figure}
\includegraphics[width=8.0cm]{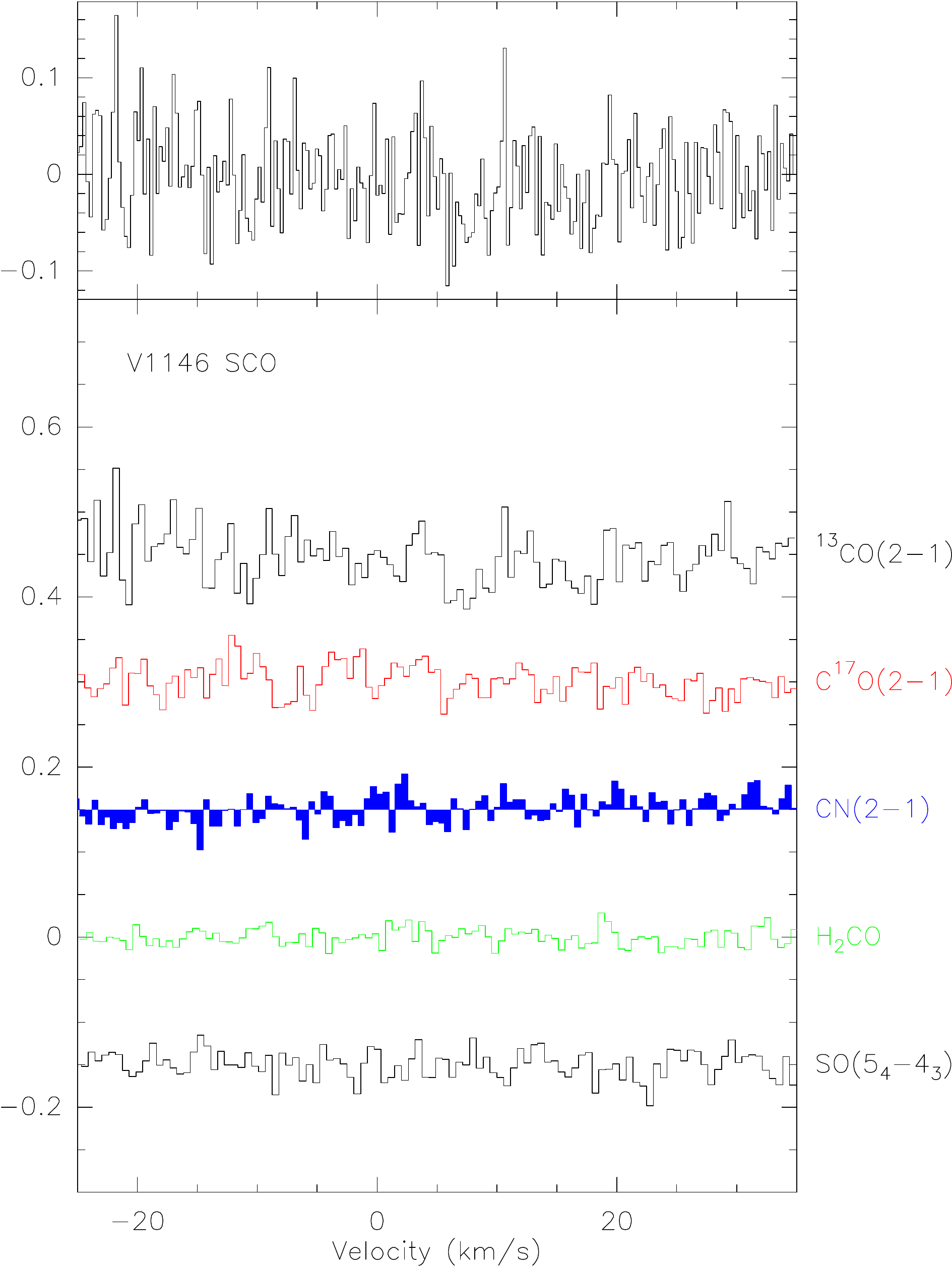}
\caption{Lines toward V1146 Sco.}
\label{fig:spe-V1146_SCO}
\end{figure}
\begin{figure}
\includegraphics[width=8.0cm]{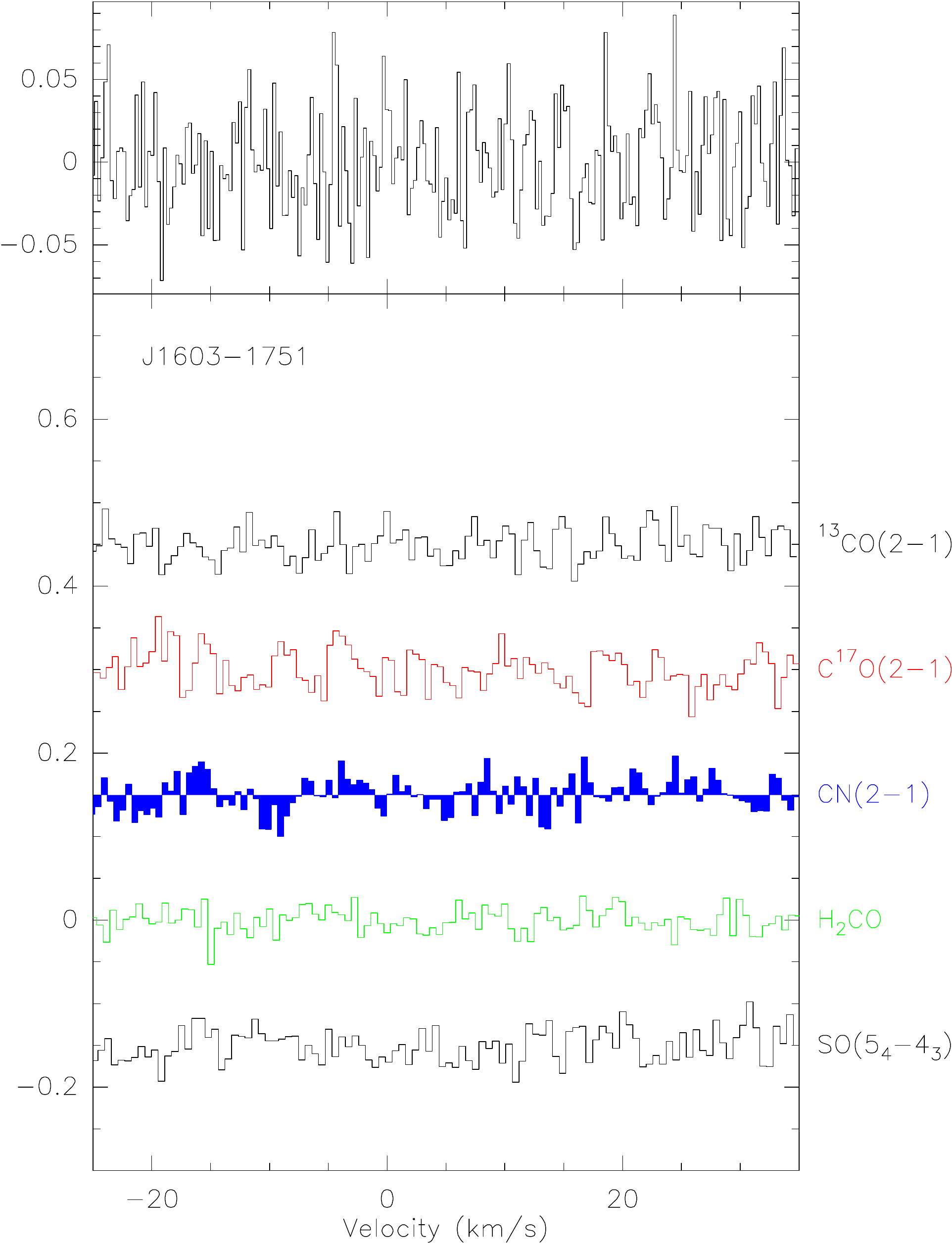}
\caption{Lines toward J1603-1751.}
\label{fig:spe-J1603-1751}
\end{figure}
\begin{figure}
\includegraphics[width=8.0cm]{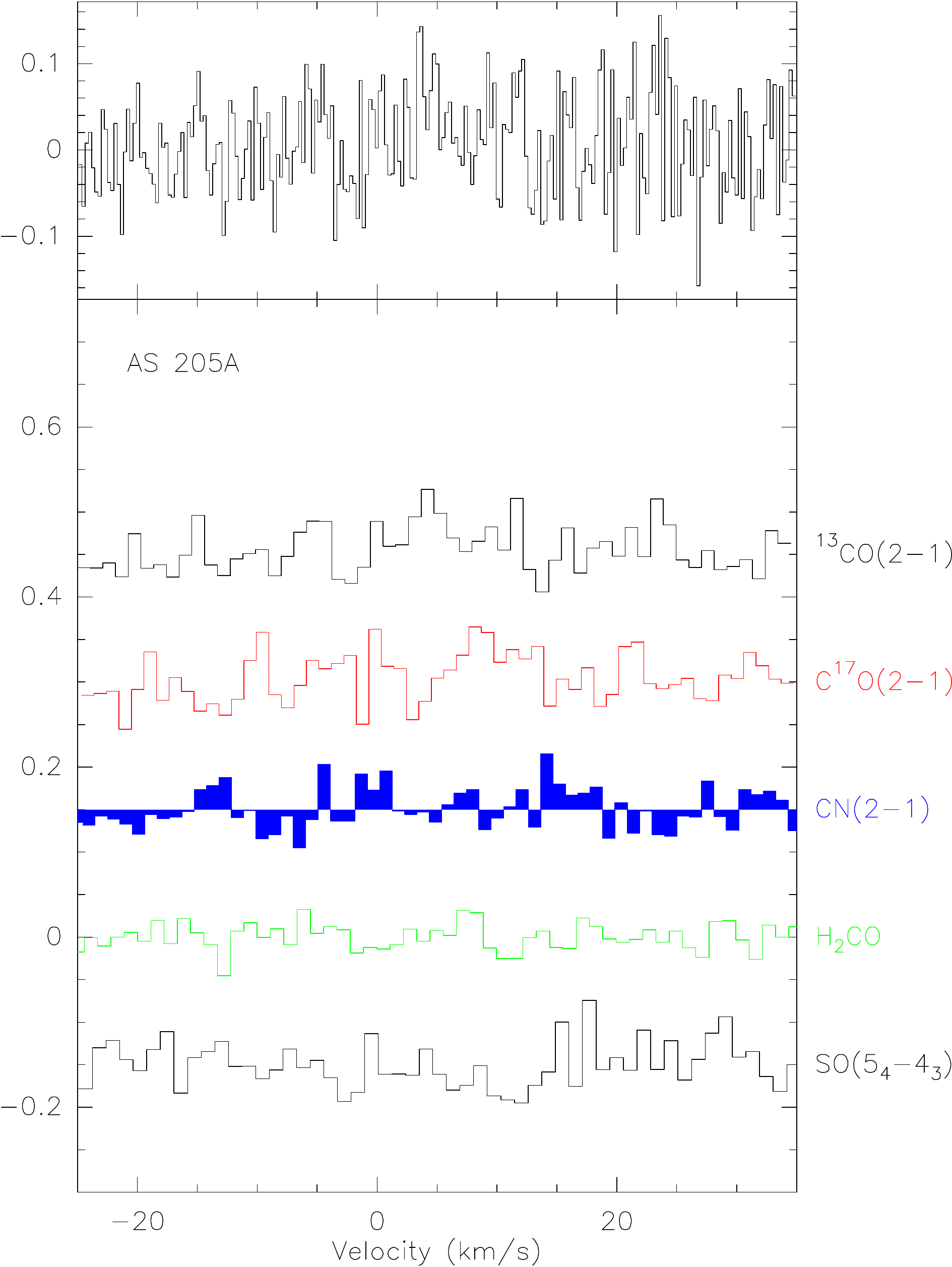}
\caption{Lines toward AS 205A.}
\label{fig:spe-AS_205A}
\end{figure}
\begin{figure}
\includegraphics[width=8.0cm]{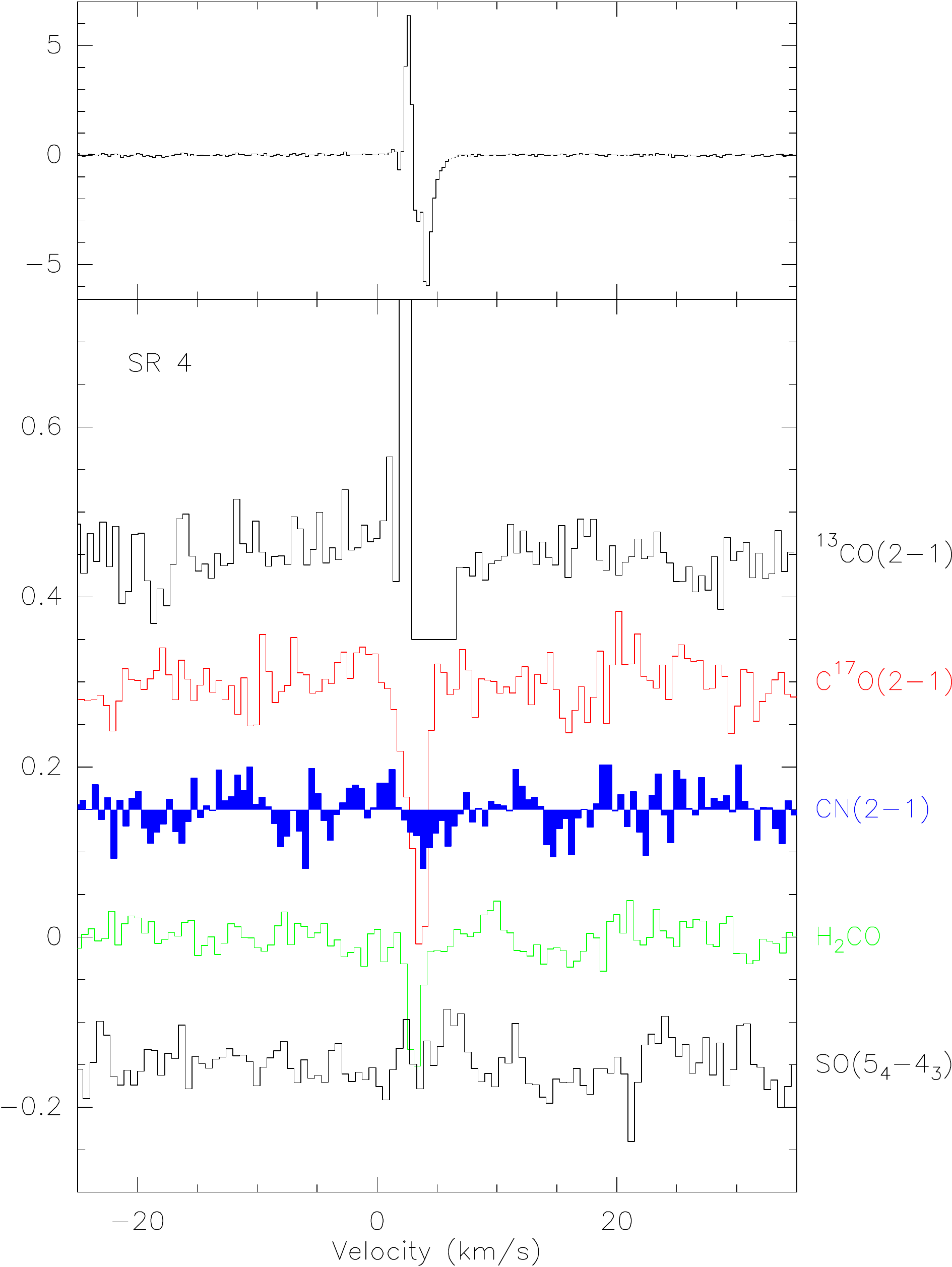}
\caption{Lines toward SR 4.}
\label{fig:spe-SR_4}
\end{figure}
\clearpage
\begin{figure}
\includegraphics[width=8.0cm]{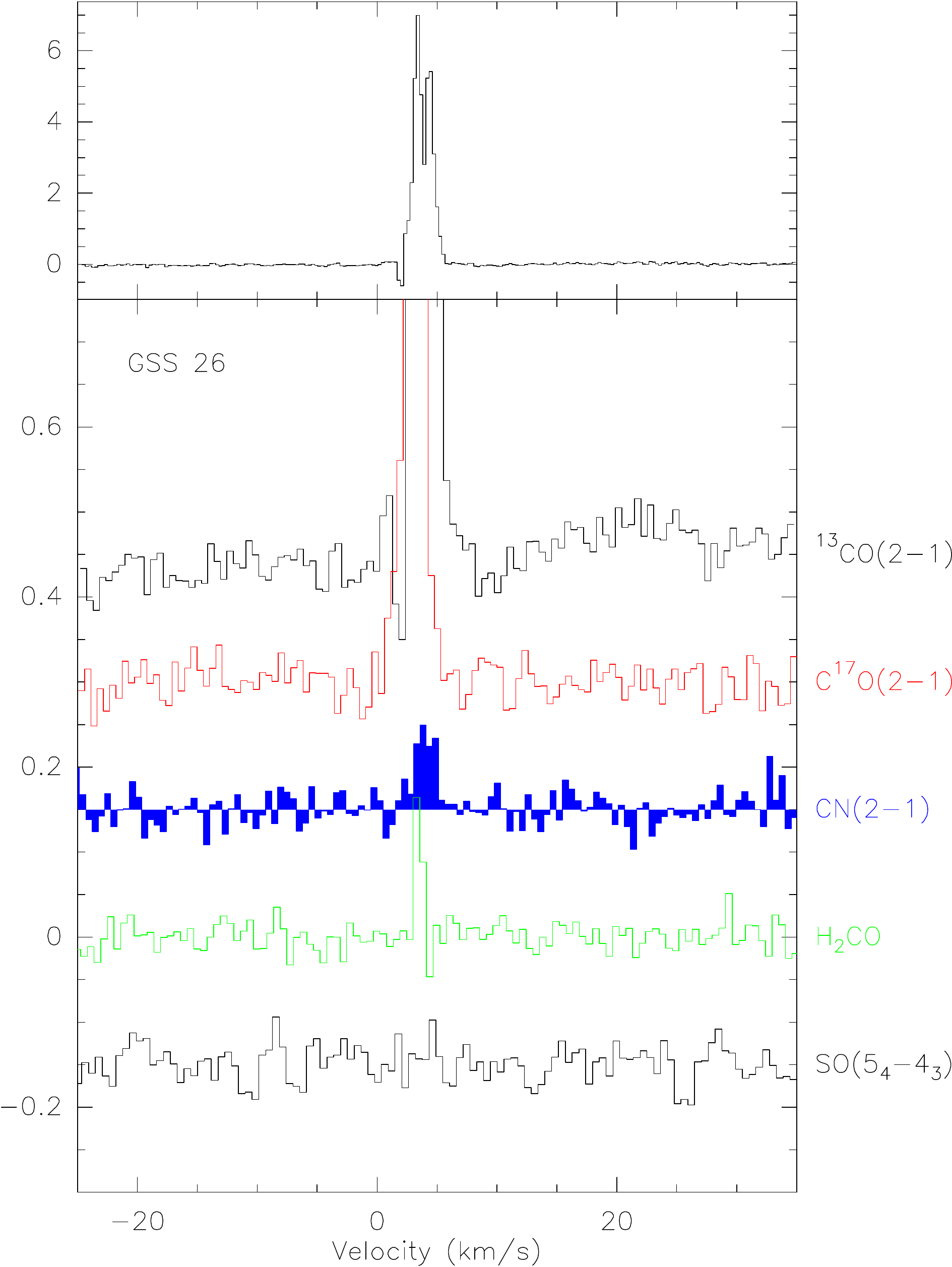}
\caption{Lines toward GSS 26.}
\label{fig:spe-GSS_26}
\end{figure}
\begin{figure}
\includegraphics[width=8.0cm]{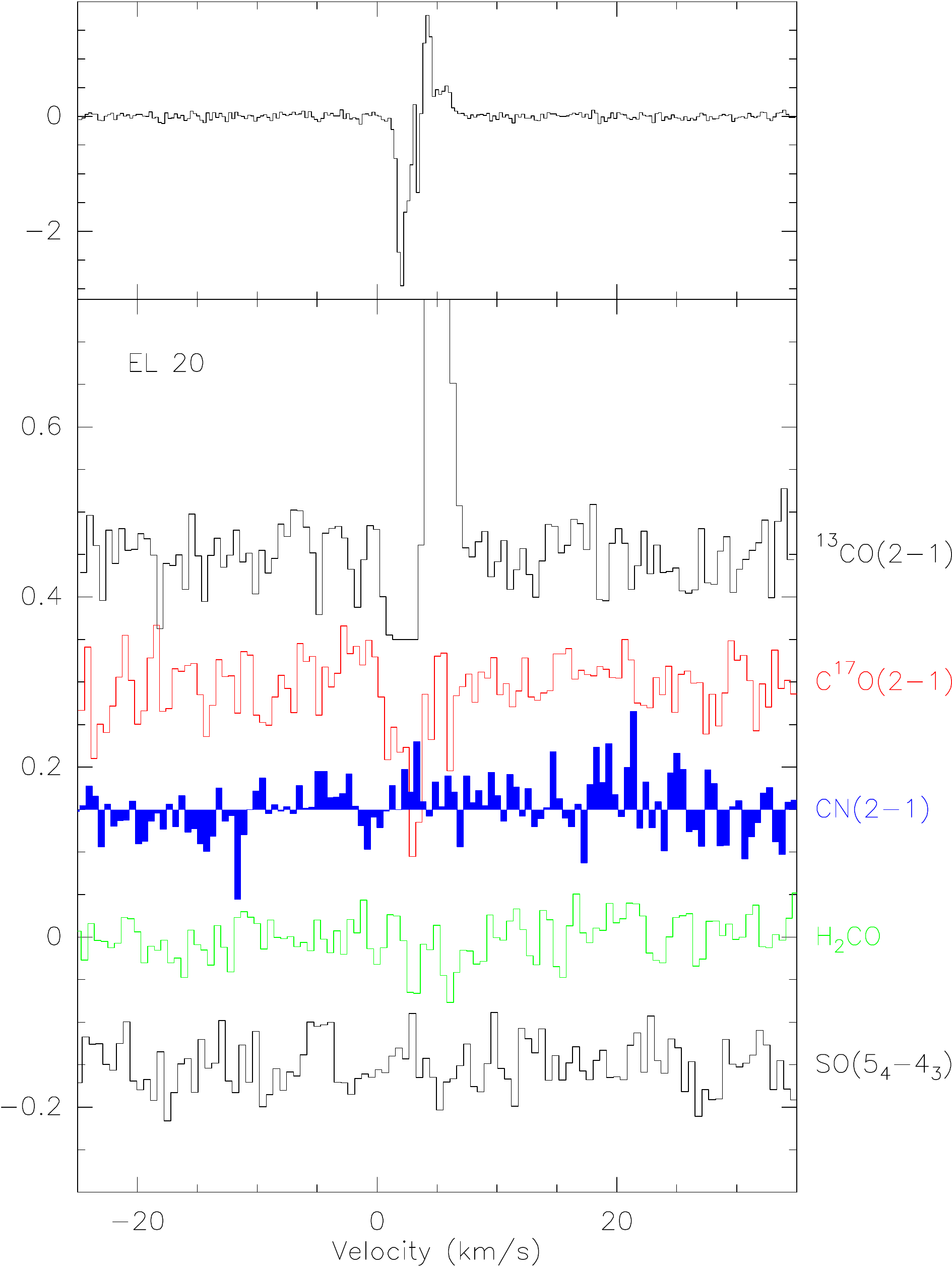}
\caption{Lines toward EL 20.}
\label{fig:spe-EL_20}
\end{figure}
\begin{figure}
\includegraphics[width=8.0cm]{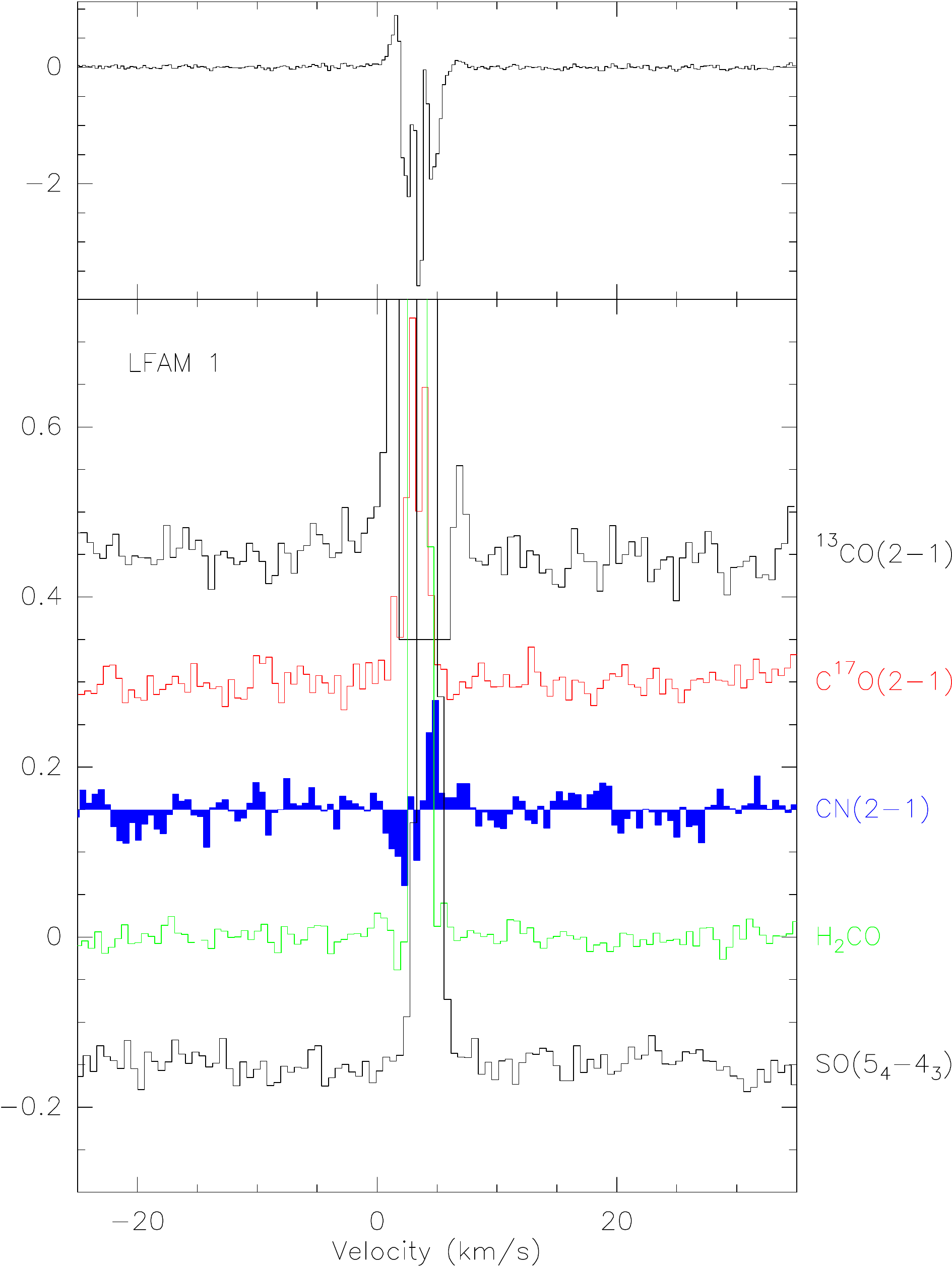}
\caption{Lines toward LFAM 1.}
\label{fig:spe-LFAM_1}
\end{figure}
\begin{figure}
\includegraphics[width=8.0cm]{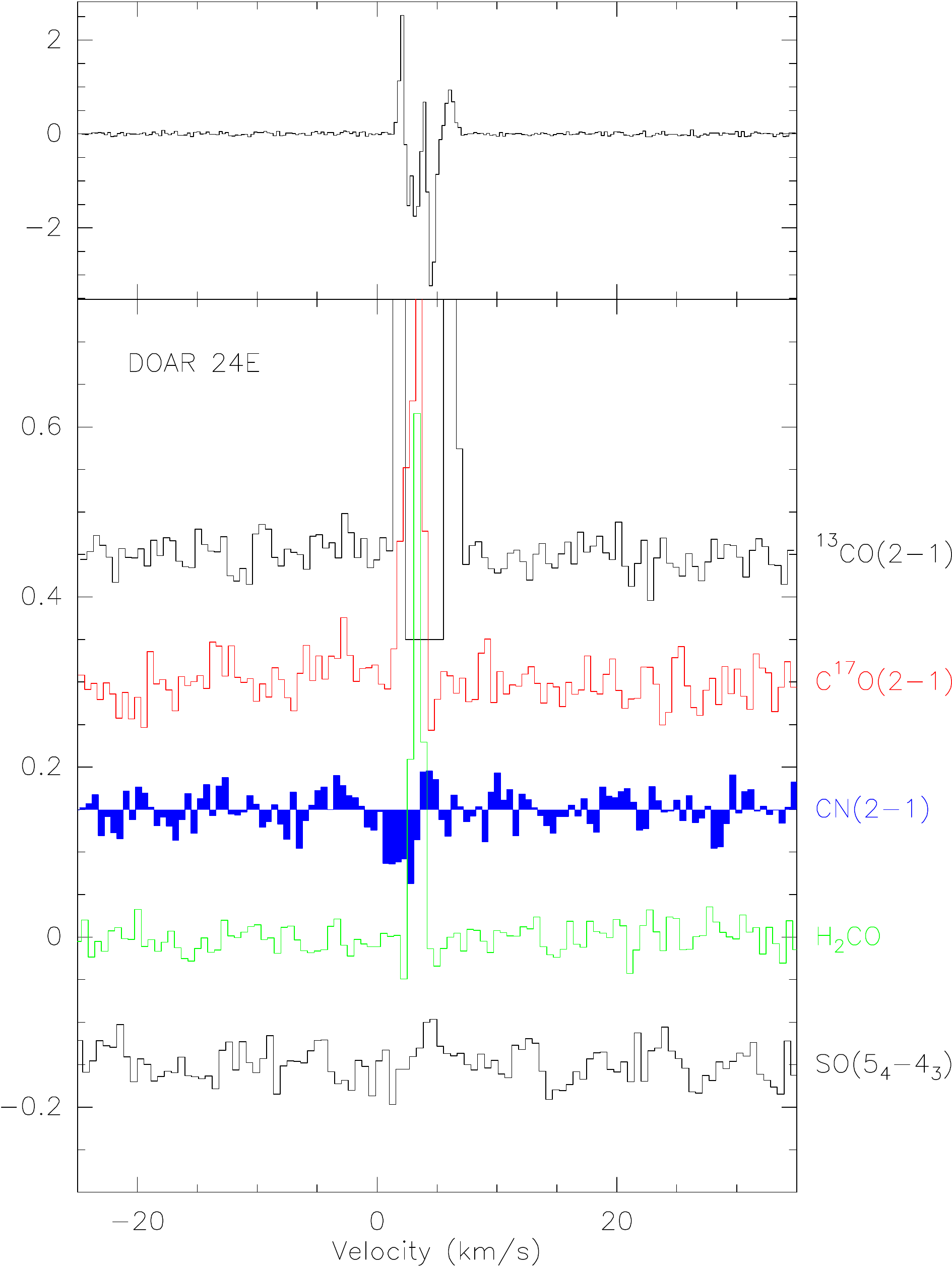}
\caption{Lines toward DoAr 24E.}
\label{fig:spe-DOAR_24E}
\end{figure}
\clearpage
\begin{figure}
\includegraphics[width=8.0cm]{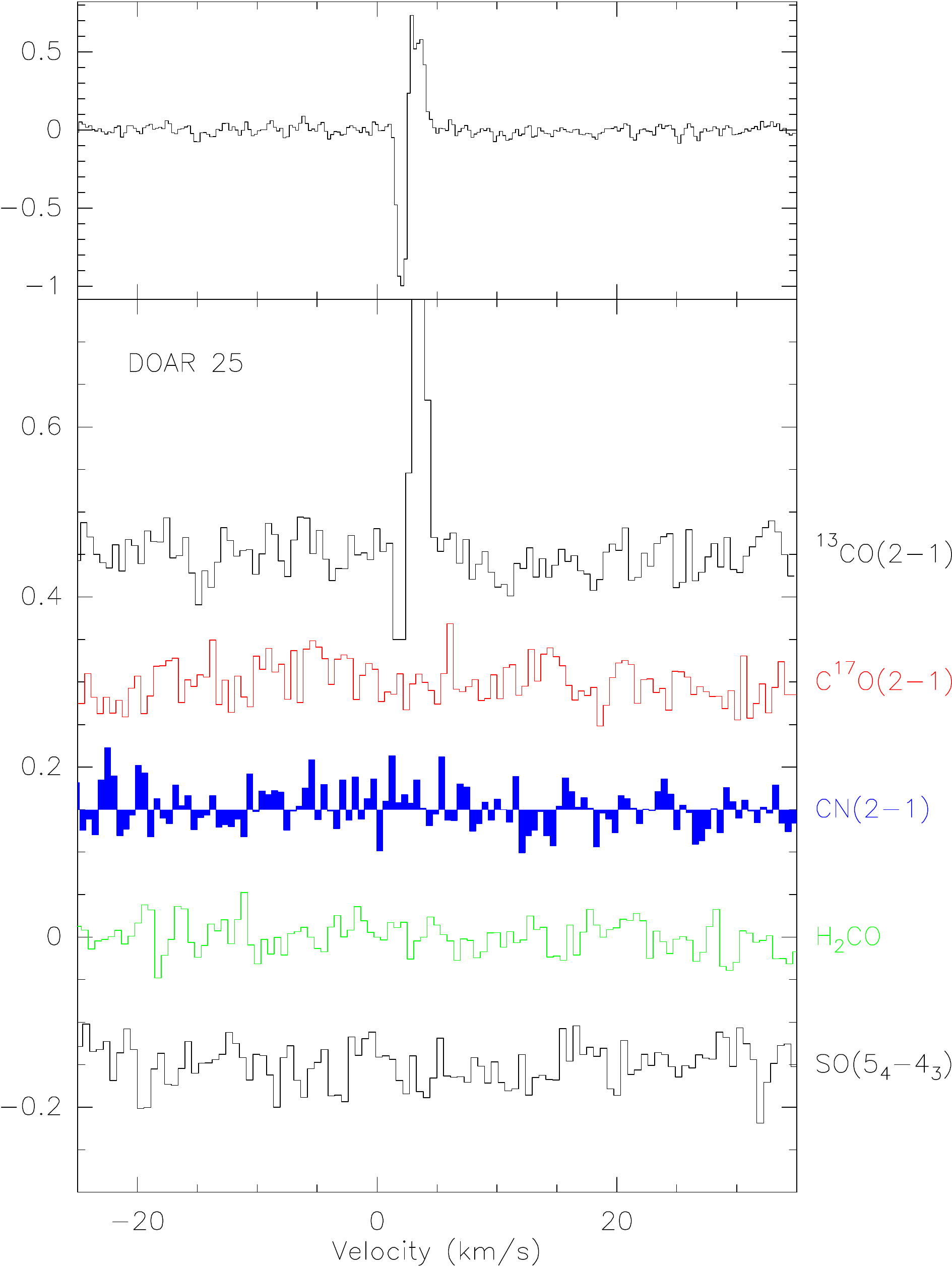}
\caption{Lines toward DoAr 25.}
\label{fig:spe-DOAR_25}
\end{figure}
\begin{figure}
\includegraphics[width=8.0cm]{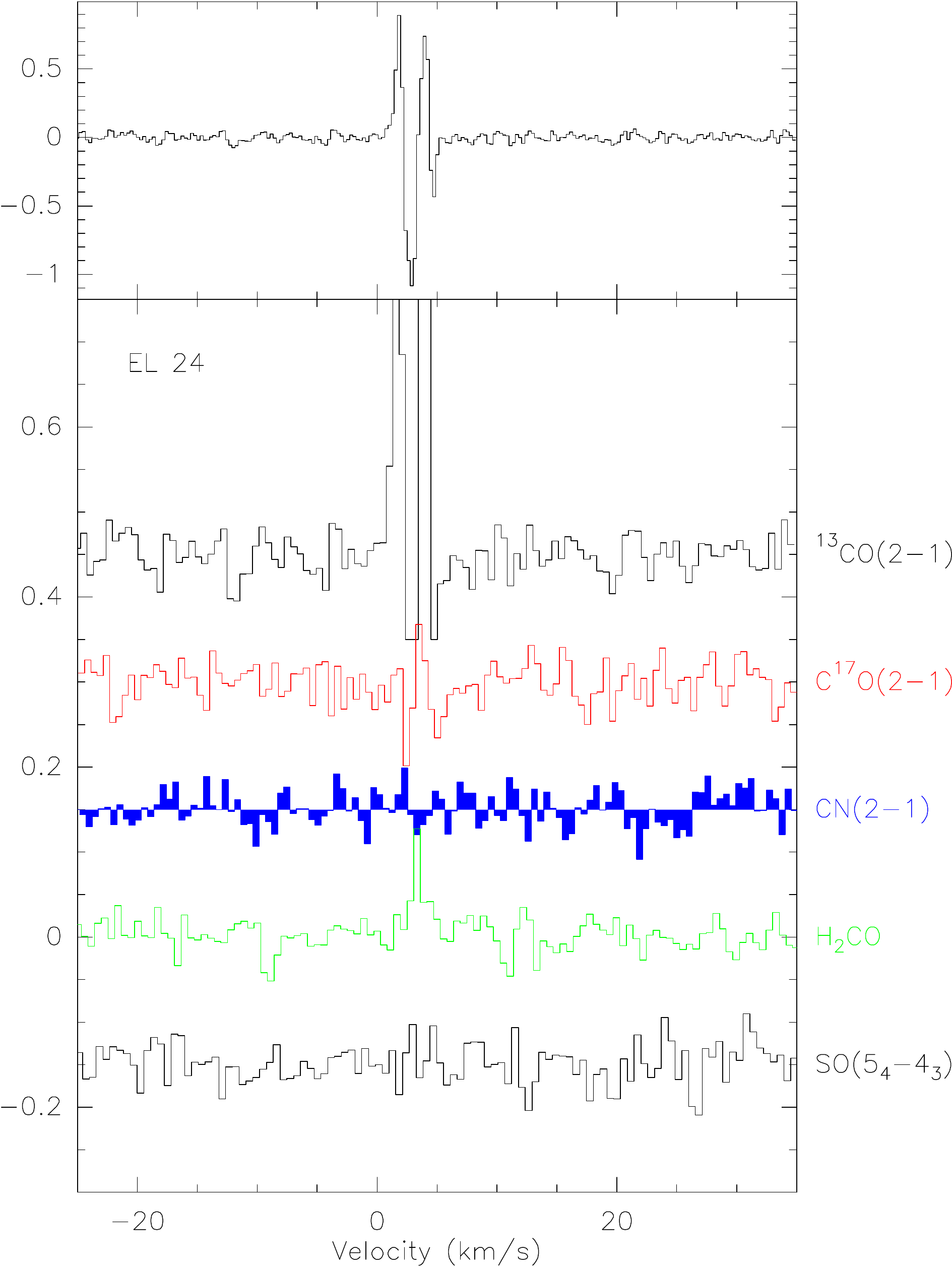}
\caption{Lines toward EL 24.}
\label{fig:spe-EL_24}
\end{figure}
\begin{figure}
\includegraphics[width=8.0cm]{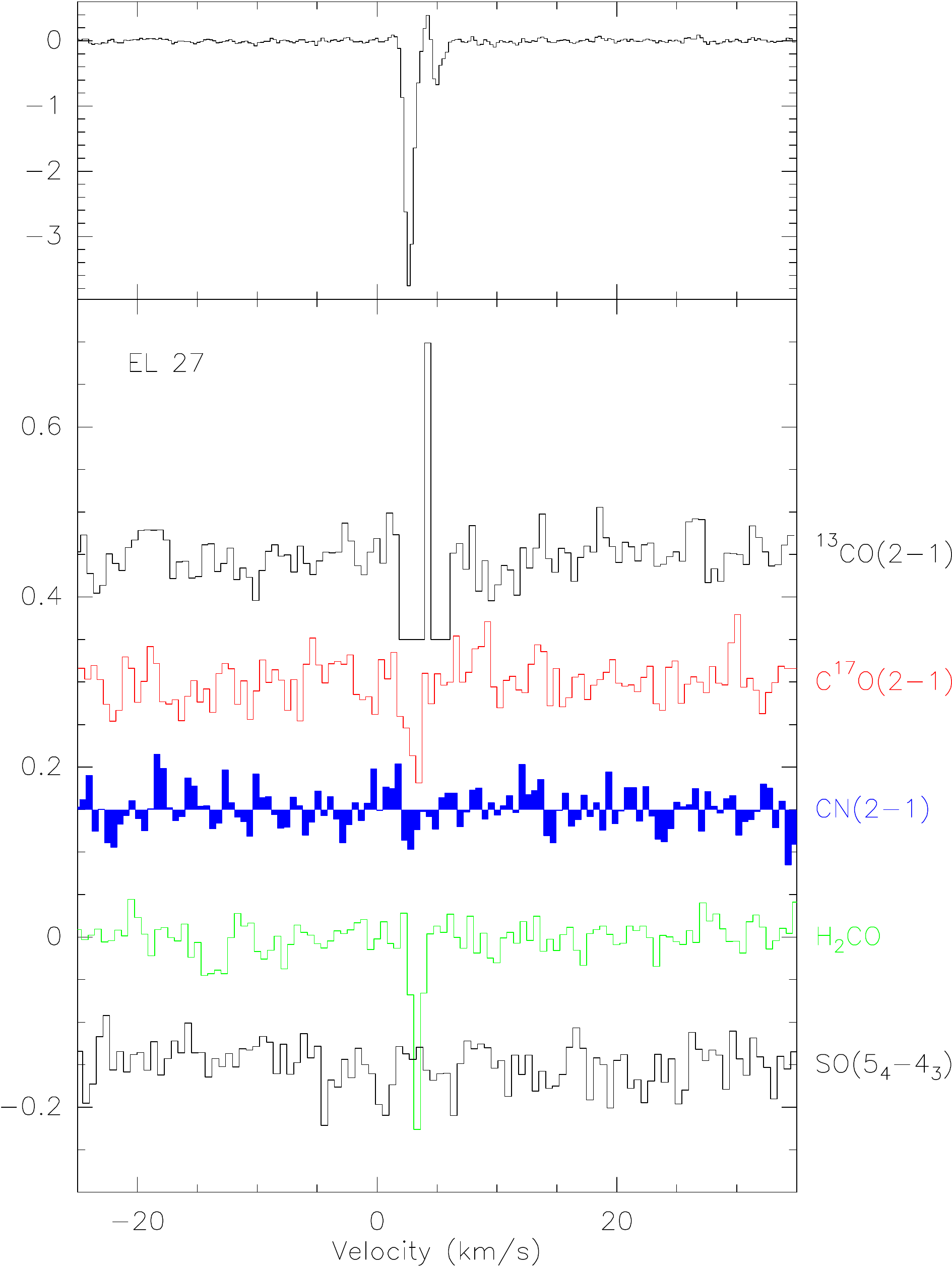}
\caption{Lines toward EL 27.}
\label{fig:spe-EL_27}
\end{figure}
\begin{figure}
\includegraphics[width=8.0cm]{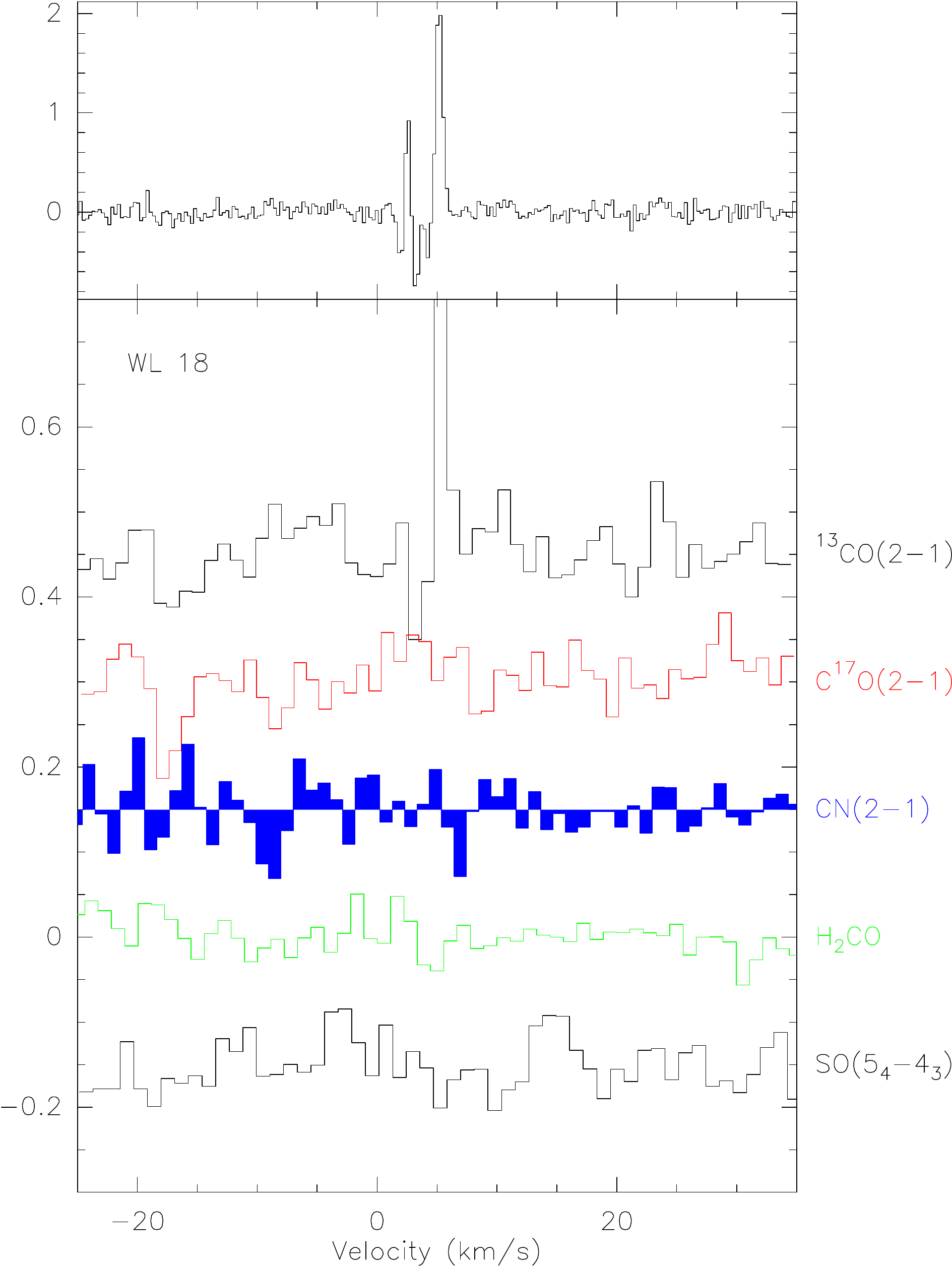}
\caption{Lines toward WL 18.}
\label{fig:spe-WL_18}
\end{figure}
\clearpage
\begin{figure}
\includegraphics[width=8.0cm]{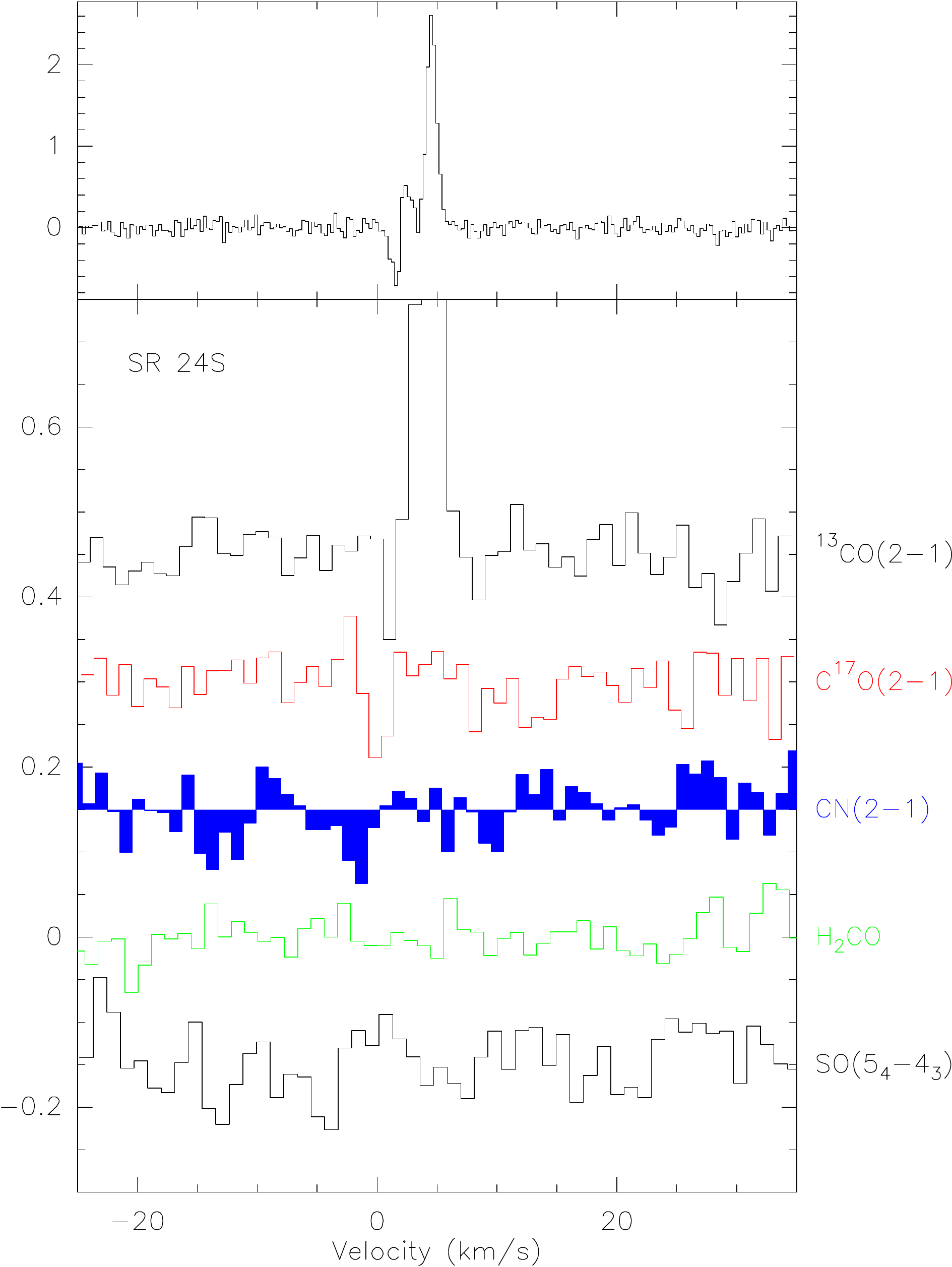}
\caption{Lines toward SR 24S.}
\label{fig:spe-SR_24S}
\end{figure}
\begin{figure}
\includegraphics[width=8.0cm]{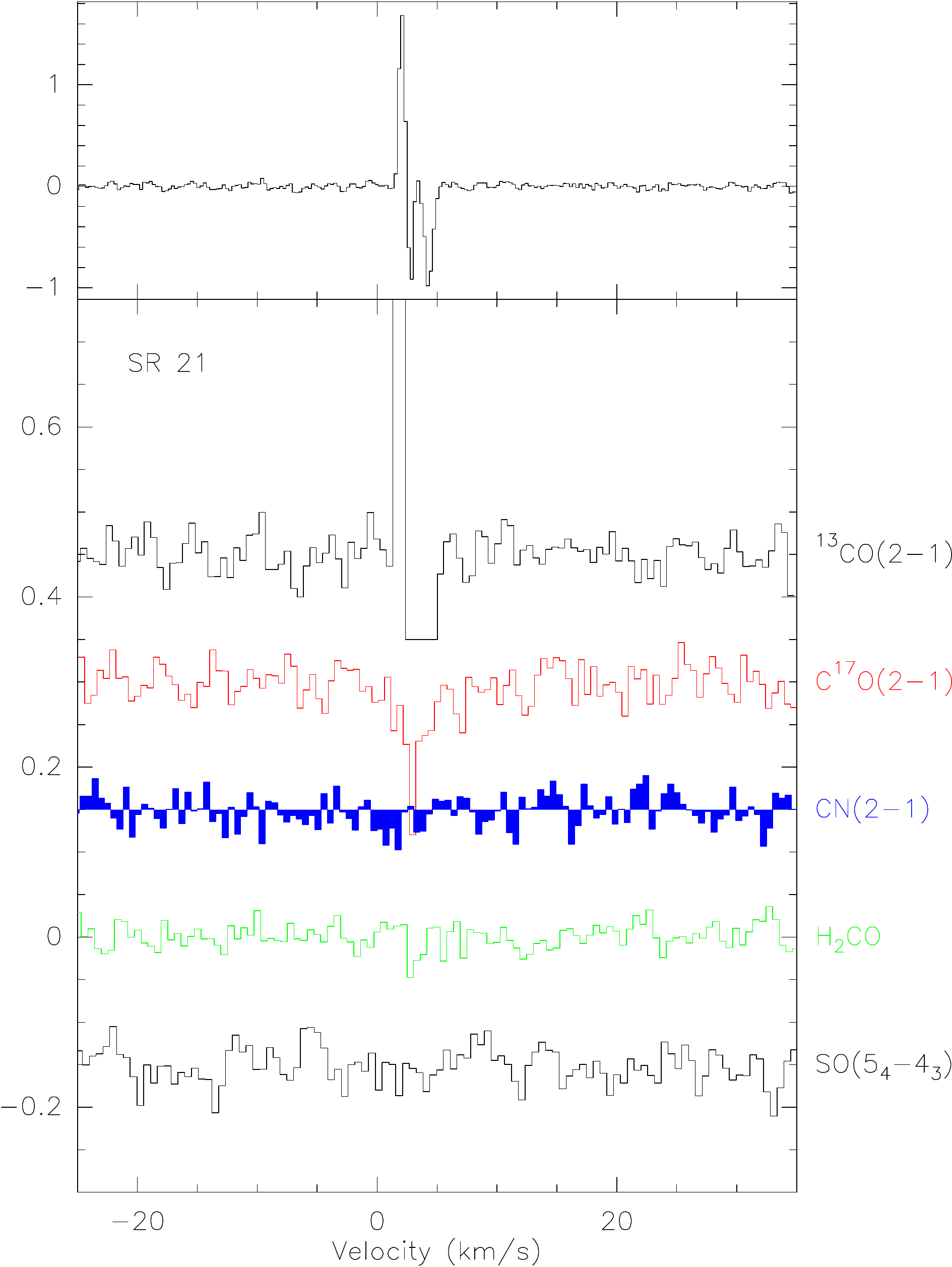}
\caption{Lines toward SR 21.}
\label{fig:spe-SR_21}
\end{figure}
\begin{figure}
\includegraphics[width=8.0cm]{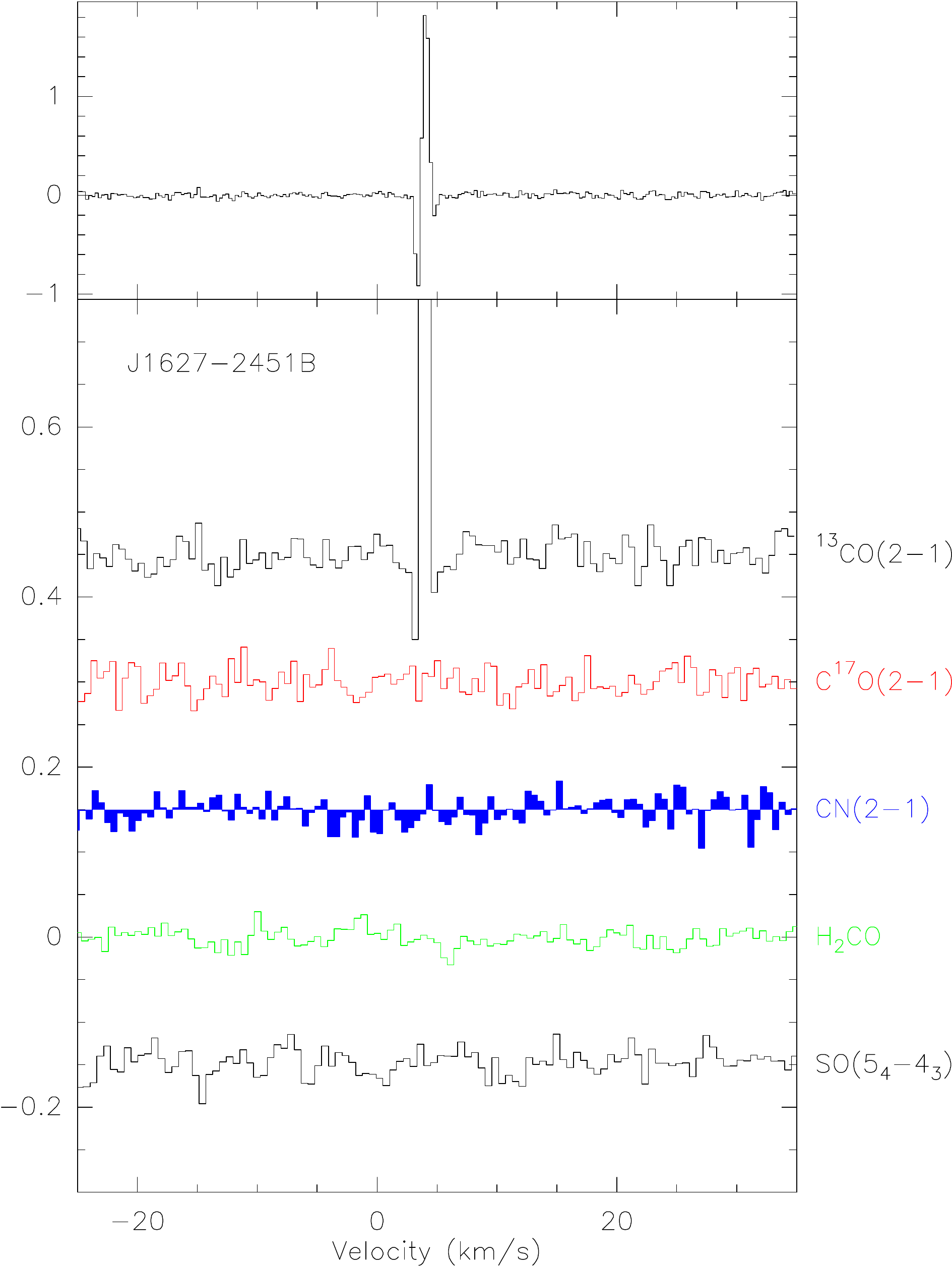}
\caption{Lines toward J1627-2451B.}
\label{fig:spe-J1627-2451B}
\end{figure}
\begin{figure}
\includegraphics[width=8.0cm]{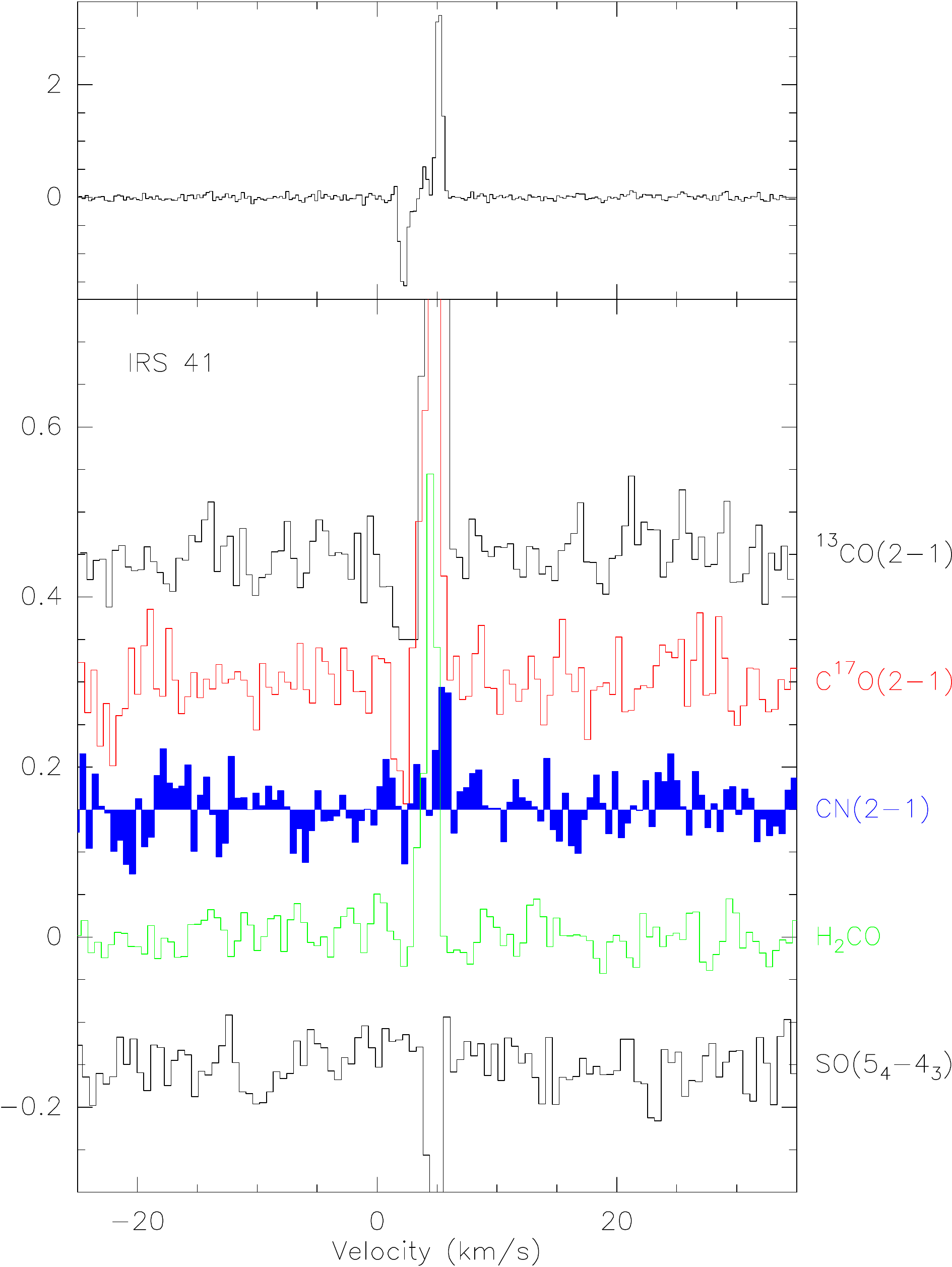}
\caption{Lines toward IRS 41.}
\label{fig:spe-IRS_41}
\end{figure}
\clearpage
\begin{figure}
\includegraphics[width=8.0cm]{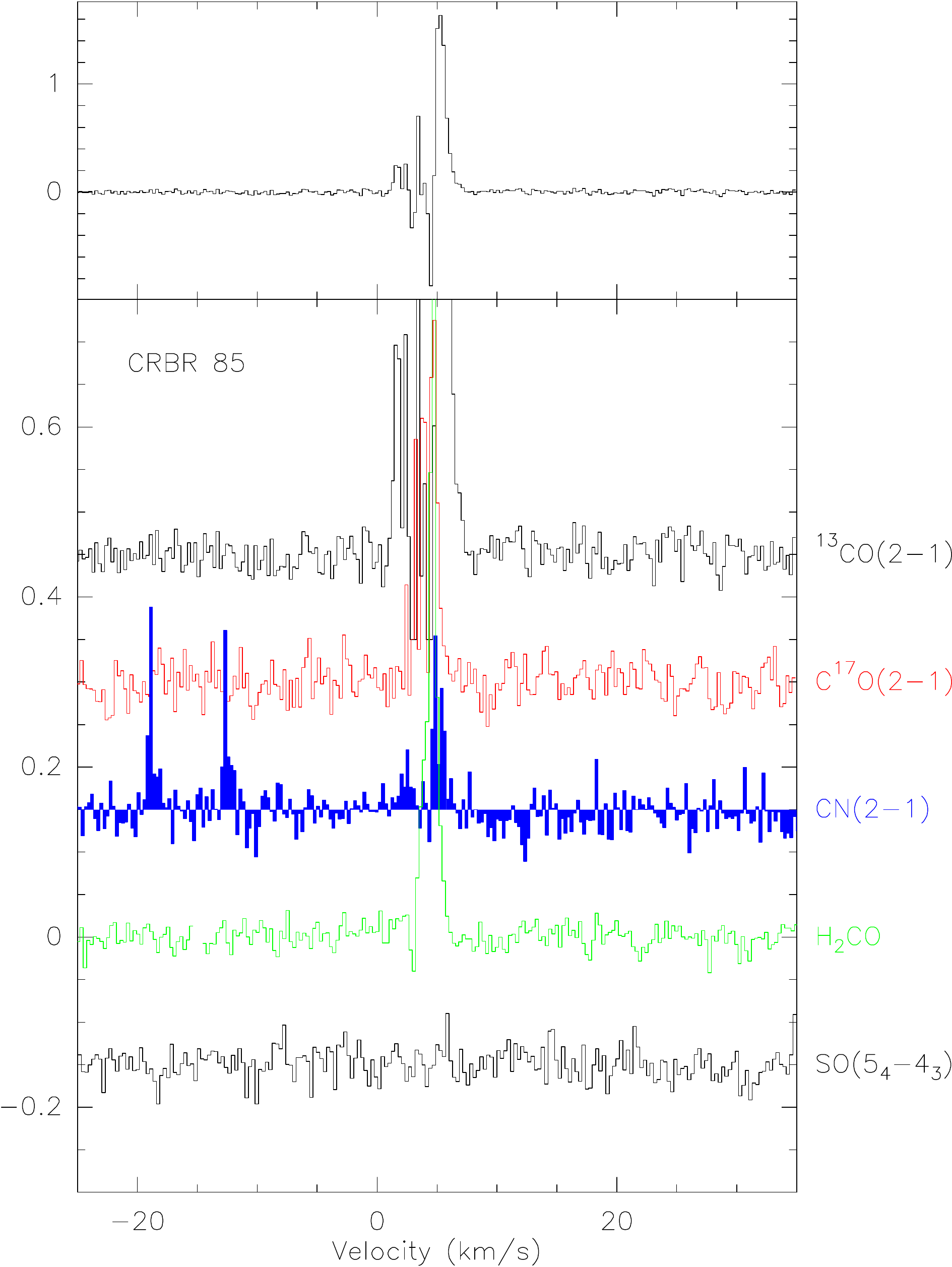}
\caption{Lines toward CRBR 85.}
\label{fig:spe-CRBR_85}
\end{figure}
\begin{figure}
\includegraphics[width=8.0cm]{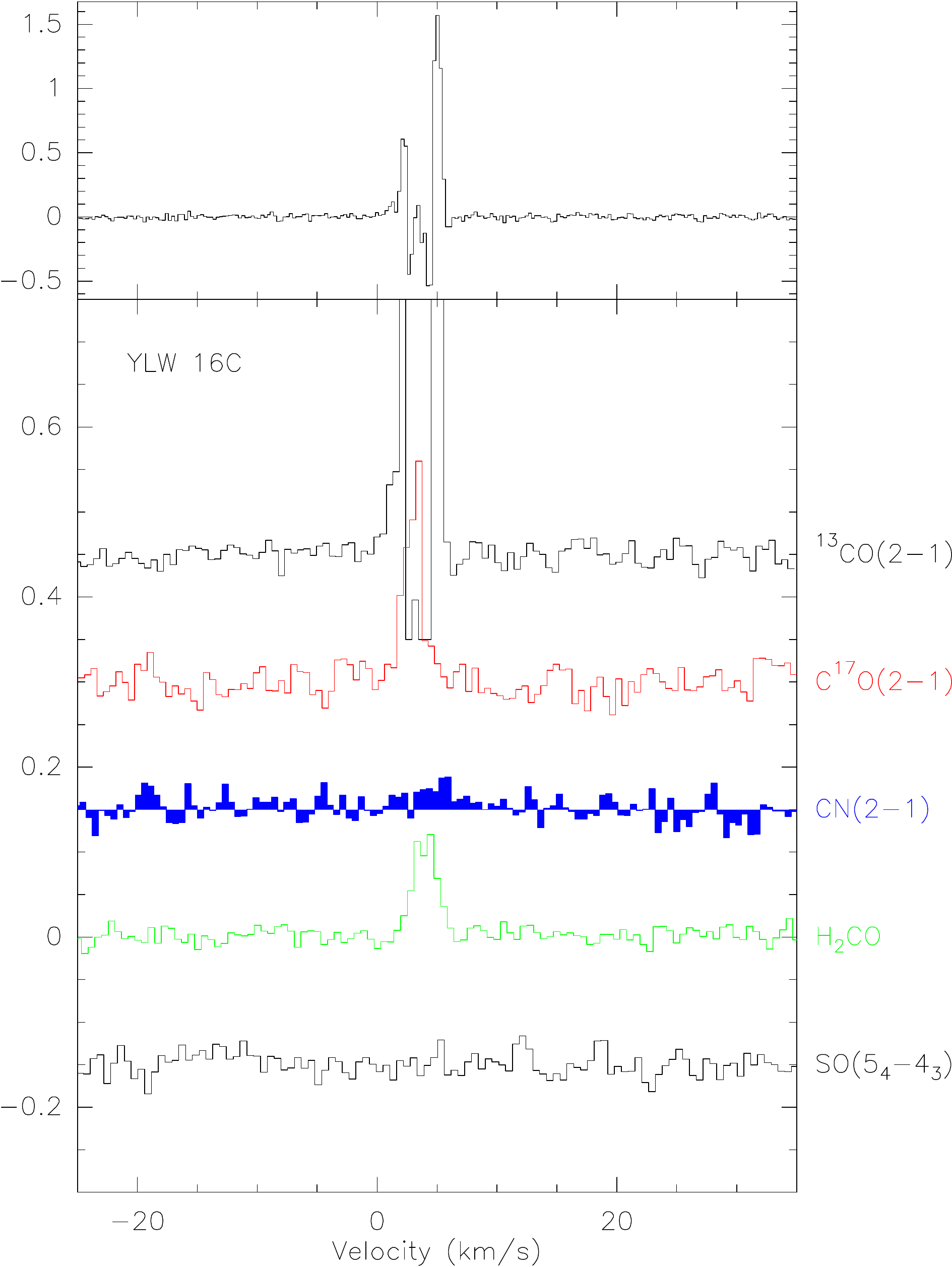}
\caption{Lines toward YLW 16c.}
\label{fig:spe-YLW_16C}
\end{figure}
\begin{figure}
\includegraphics[width=8.0cm]{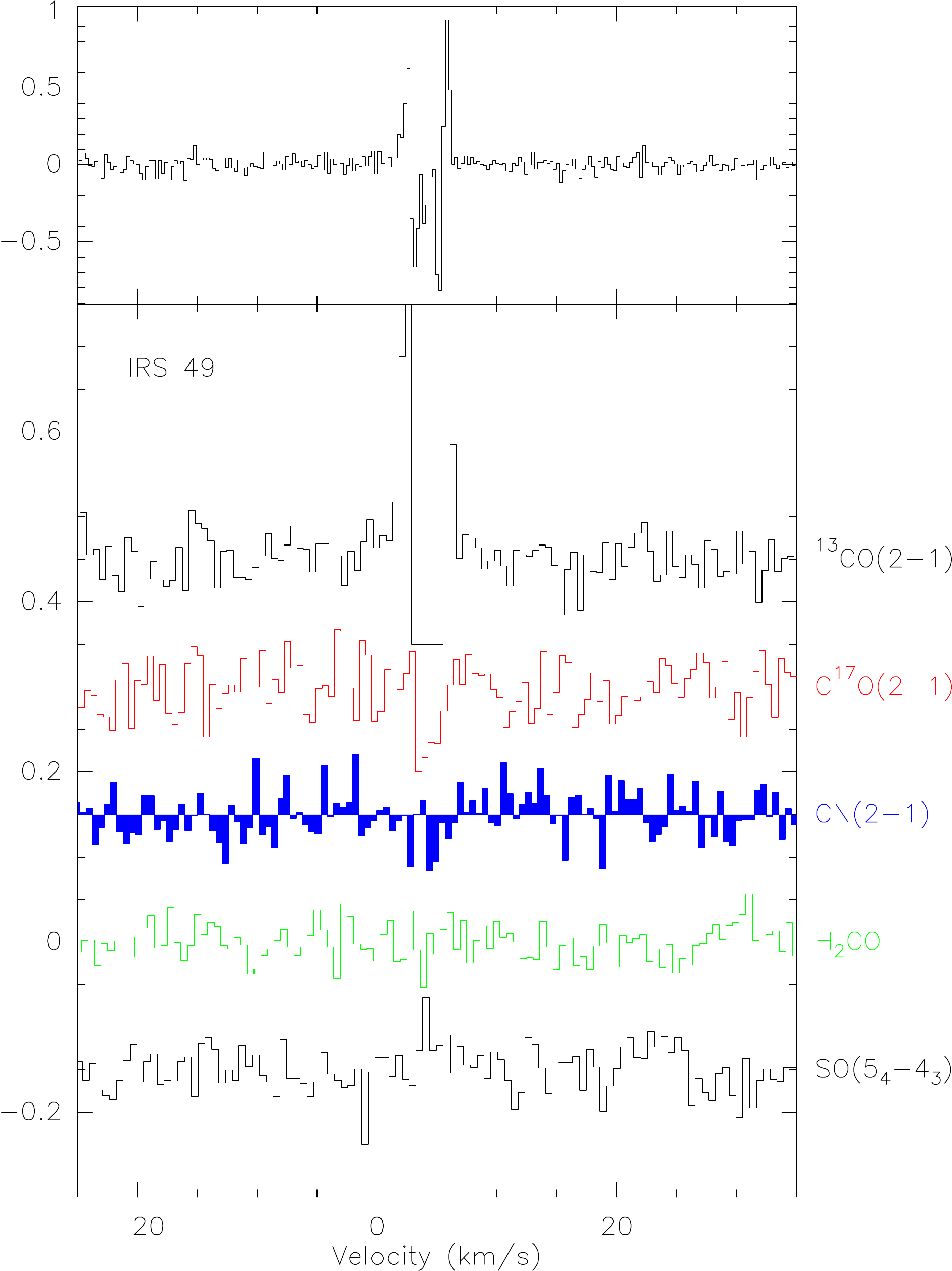}
\caption{Lines toward IRS 49.}
\label{fig:spe-IRS_49}
\end{figure}
\begin{figure}
\includegraphics[width=8.0cm]{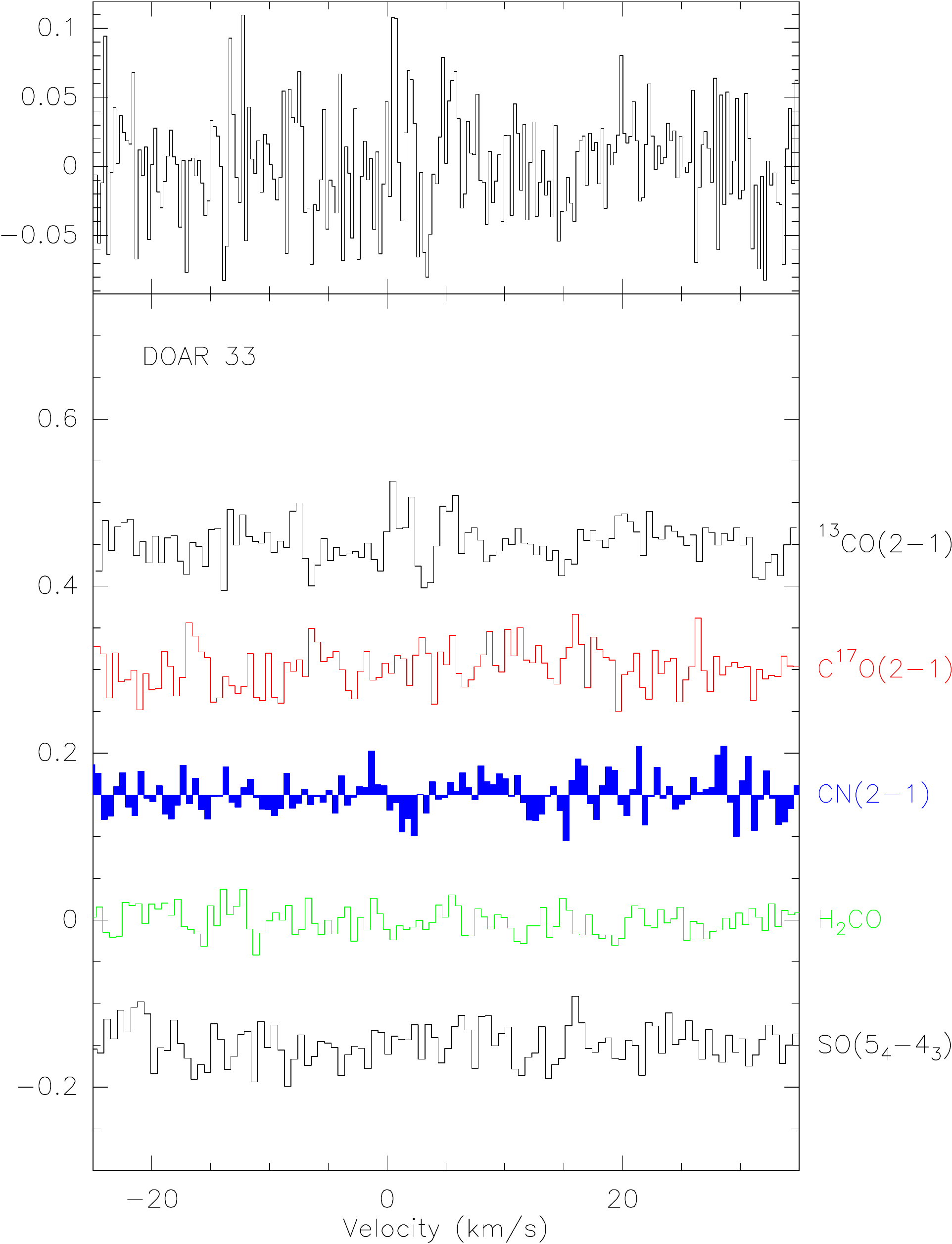}
\caption{Lines toward DoAr 33.}
\label{fig:spe-DOAR_33}
\end{figure}
\clearpage
\begin{figure}
\includegraphics[width=8.0cm]{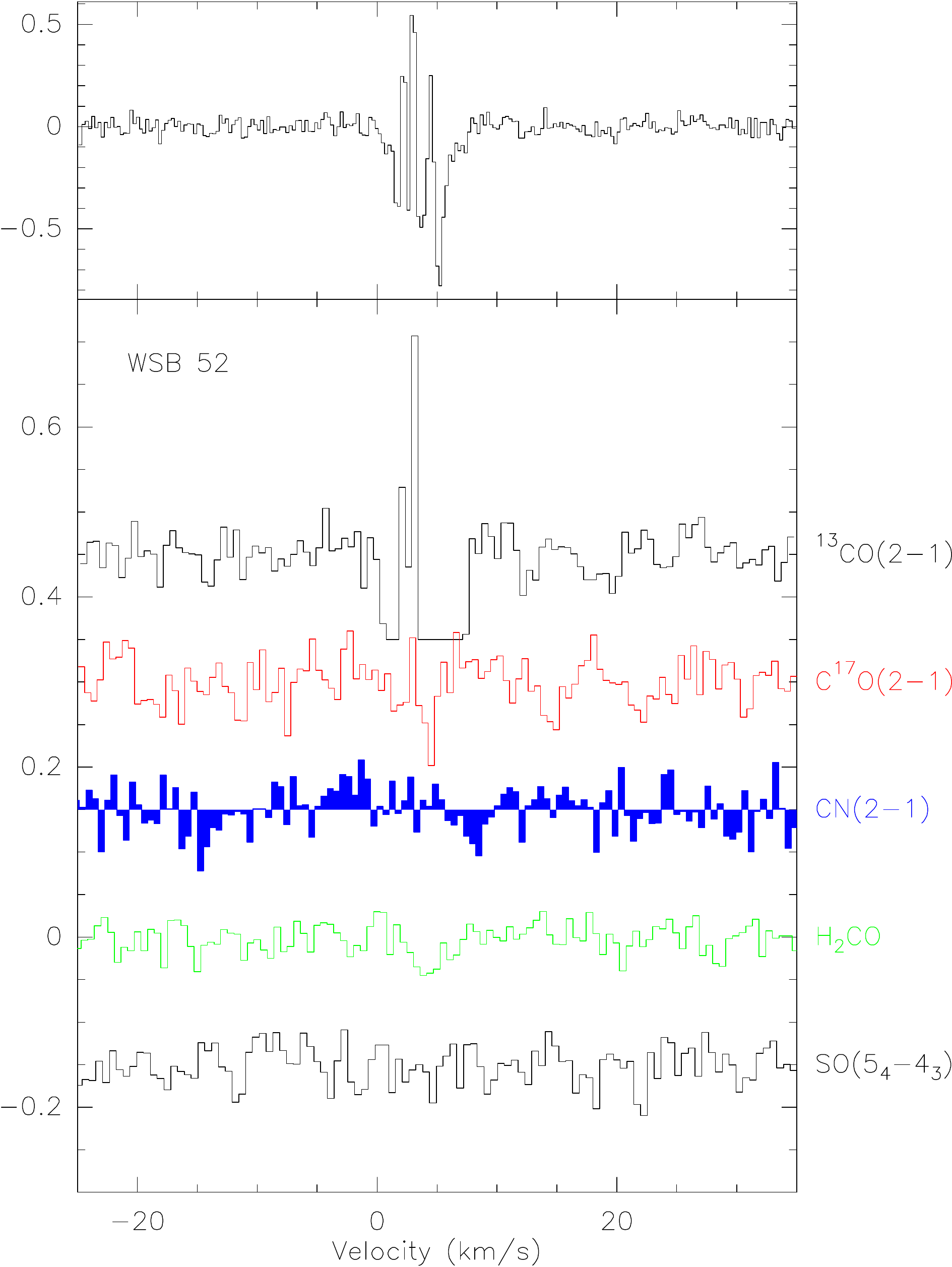}
\caption{Lines toward WSB 52.}
\label{fig:spe-WSB_52}
\end{figure}
\begin{figure}
\includegraphics[width=8.0cm]{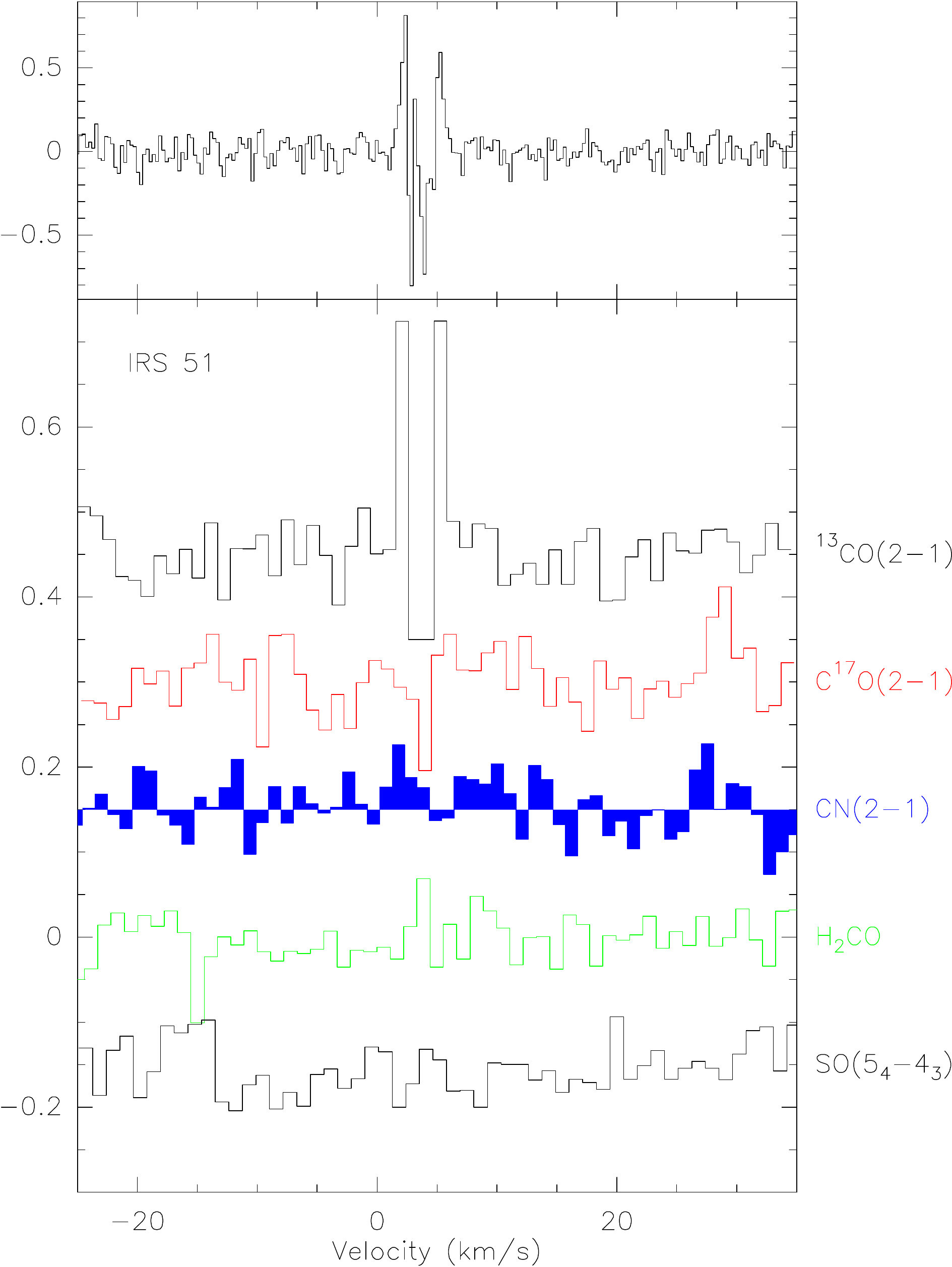}
\caption{Lines toward IRS 51.}
\label{fig:spe-IRS_51}
\end{figure}
\begin{figure}
\includegraphics[width=8.0cm]{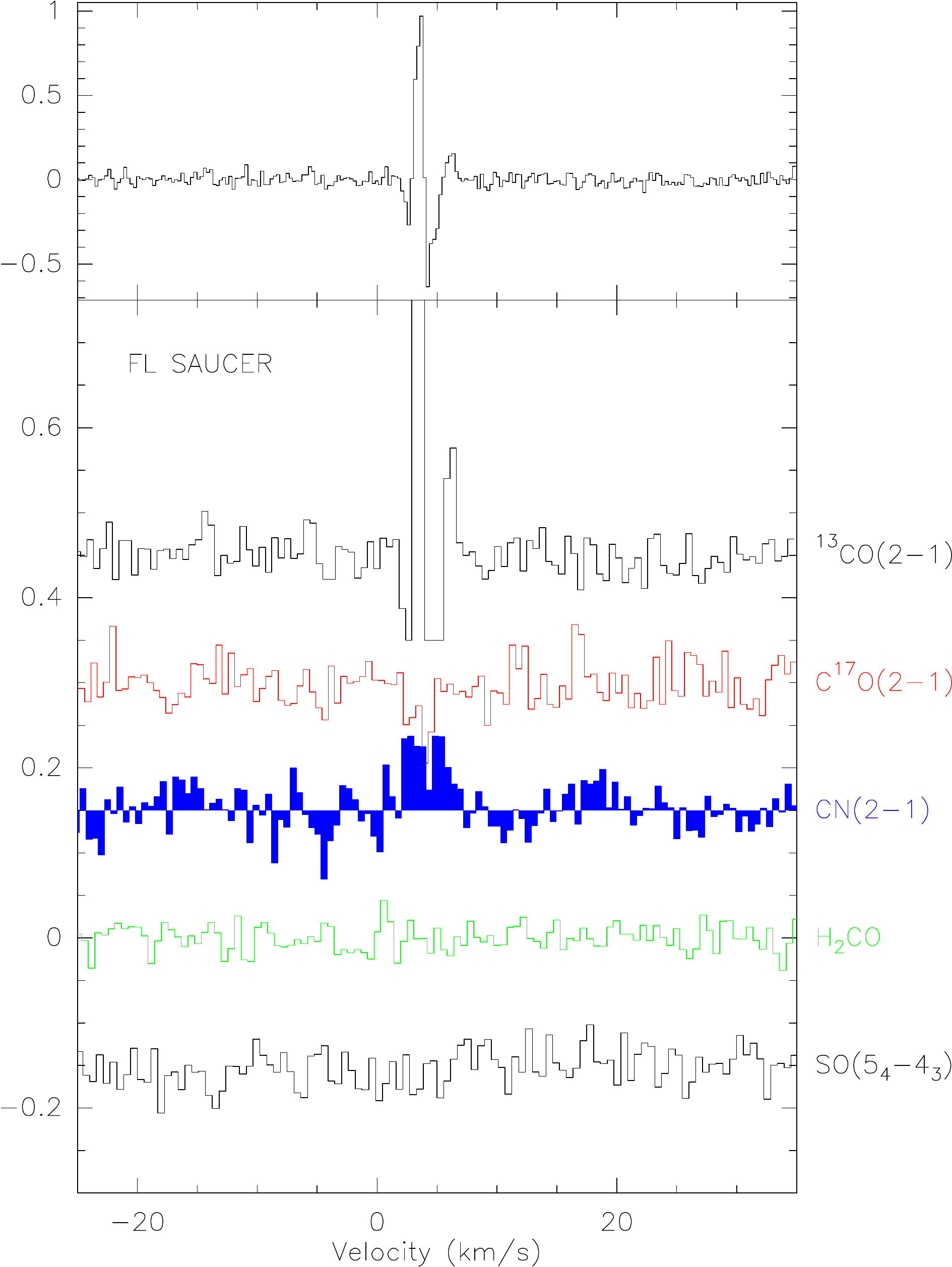}
\caption{Lines toward Flying Saucer.}
\label{fig:spe-FL_SAUCER}
\end{figure}
\begin{figure}
\includegraphics[width=8.0cm]{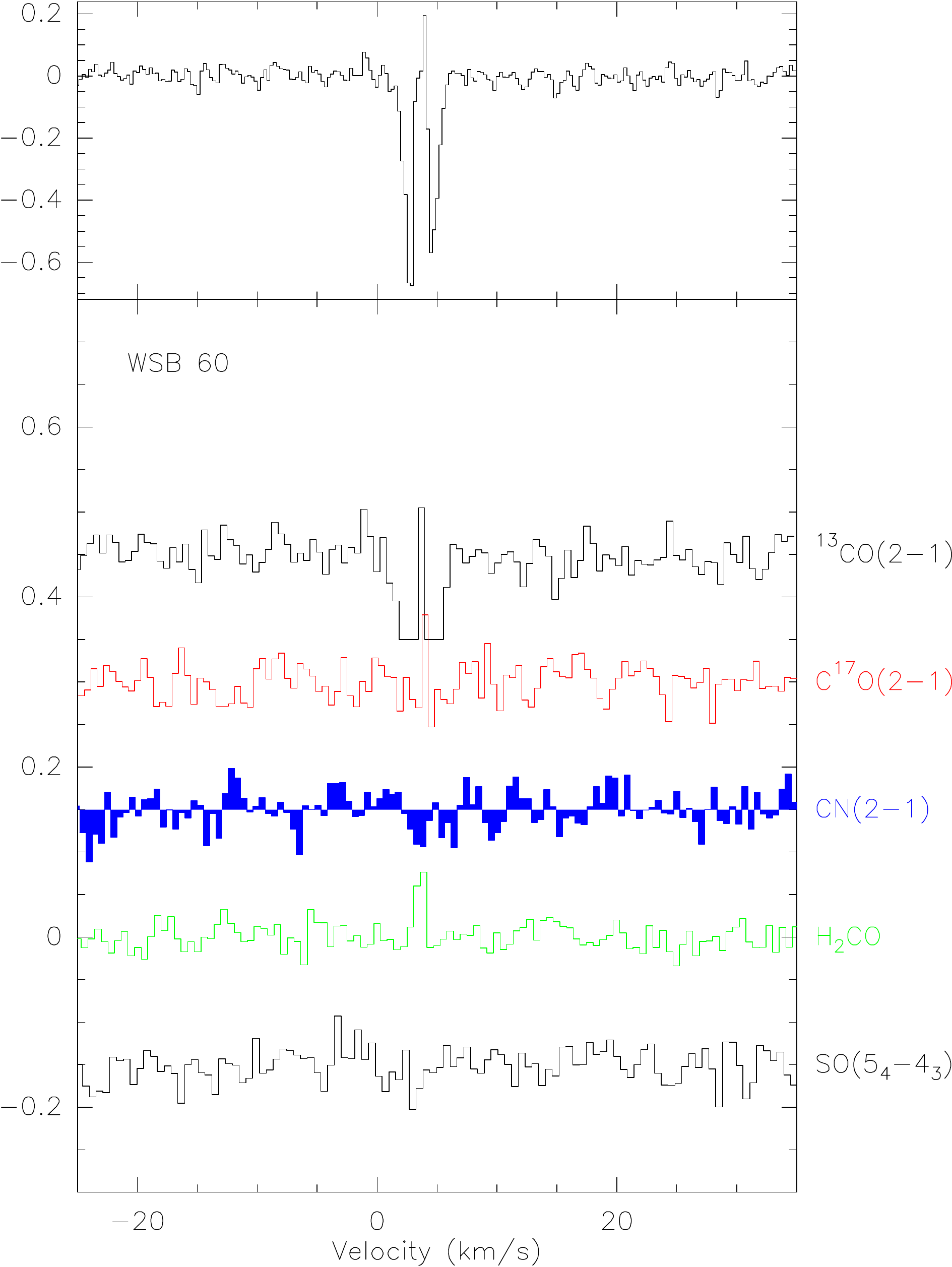}
\caption{Lines toward WSB 60.}
\label{fig:spe-WSB_60}
\end{figure}
\clearpage
\begin{figure}
\includegraphics[width=8.0cm]{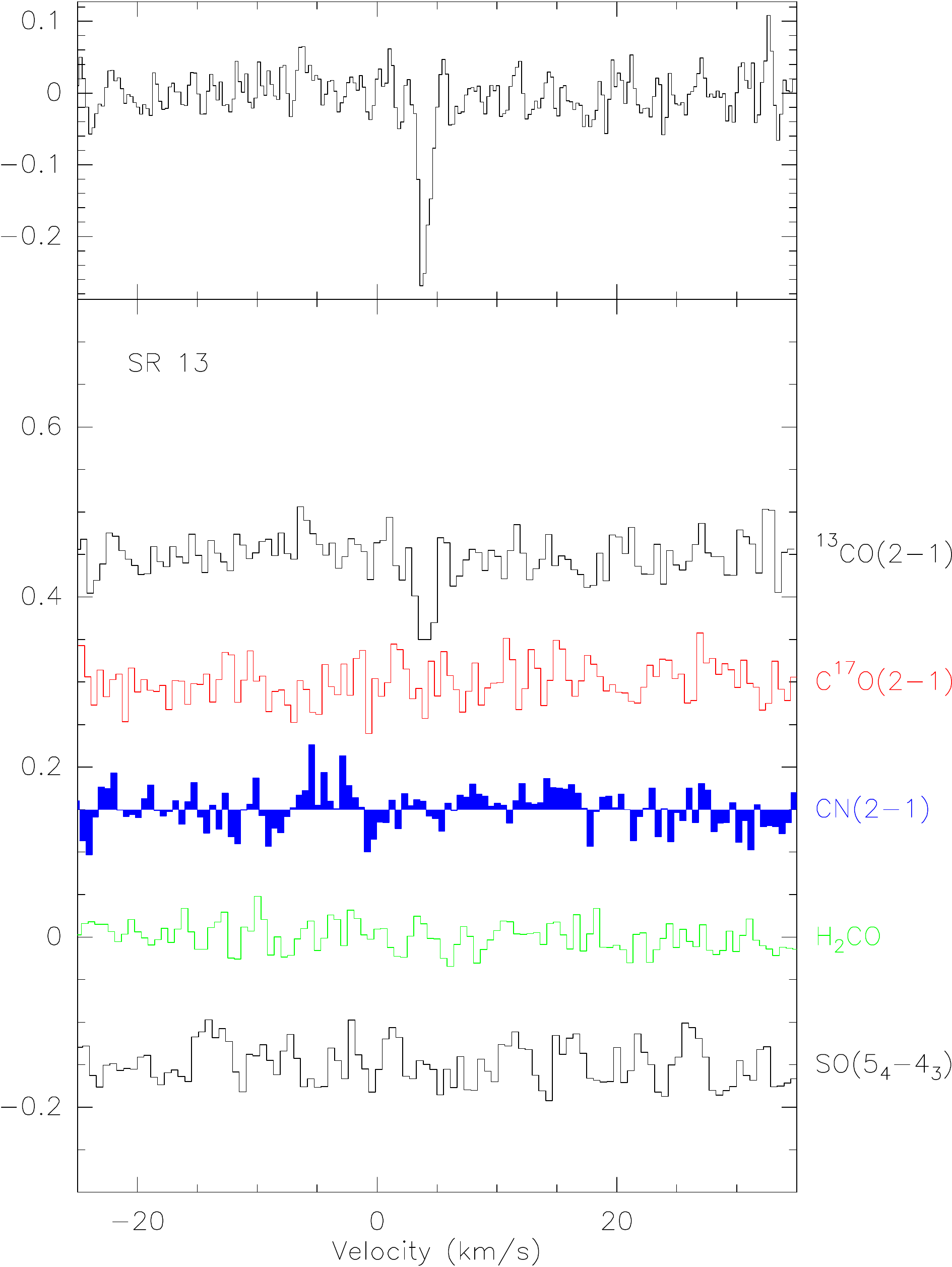}
\caption{Lines toward SR 13.}
\label{fig:spe-SR_13}
\end{figure}
\begin{figure}
\includegraphics[width=8.0cm]{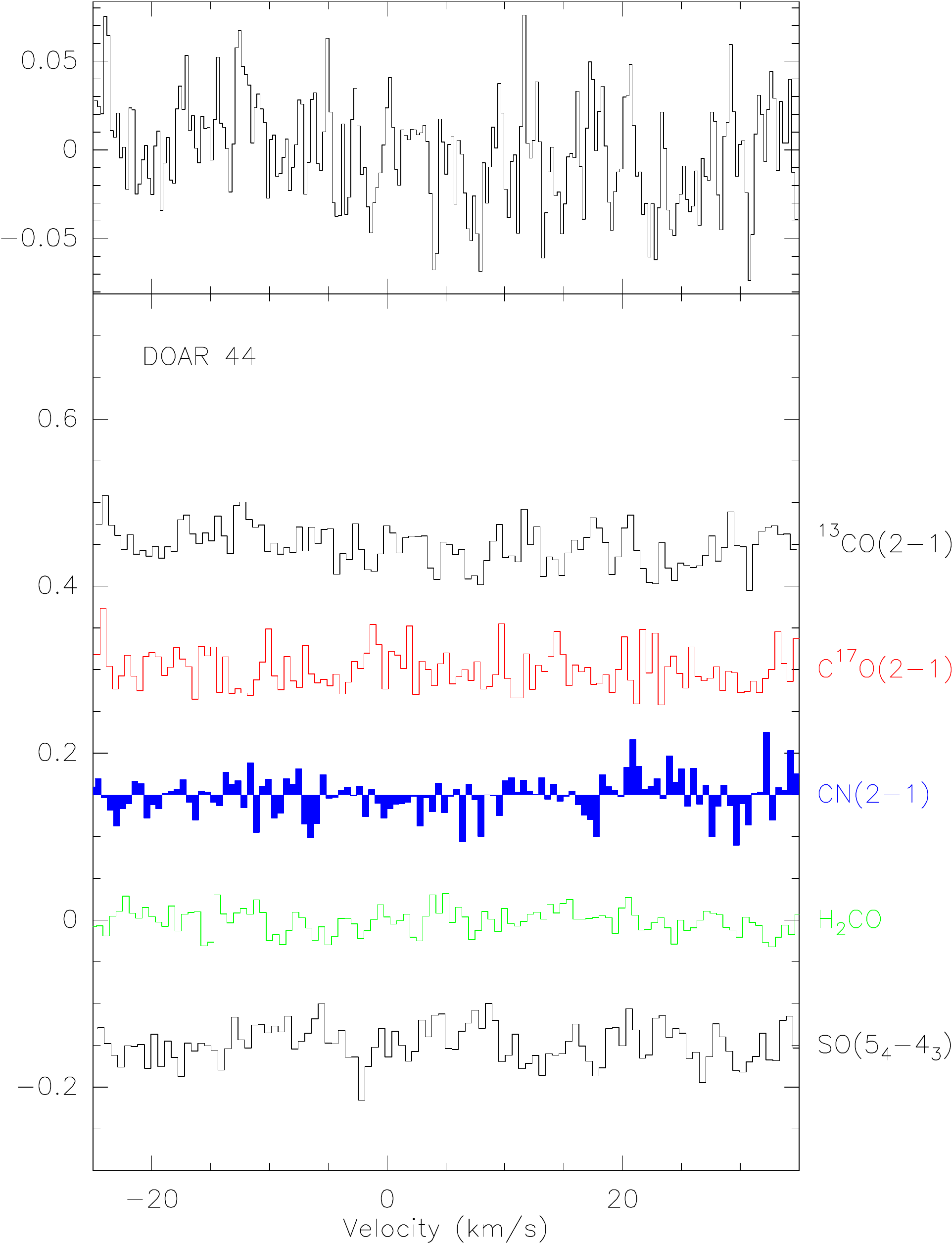}
\caption{Lines toward DoAr 44.}
\label{fig:spe-DOAR_44}
\end{figure}
\begin{figure}
\includegraphics[width=8.0cm]{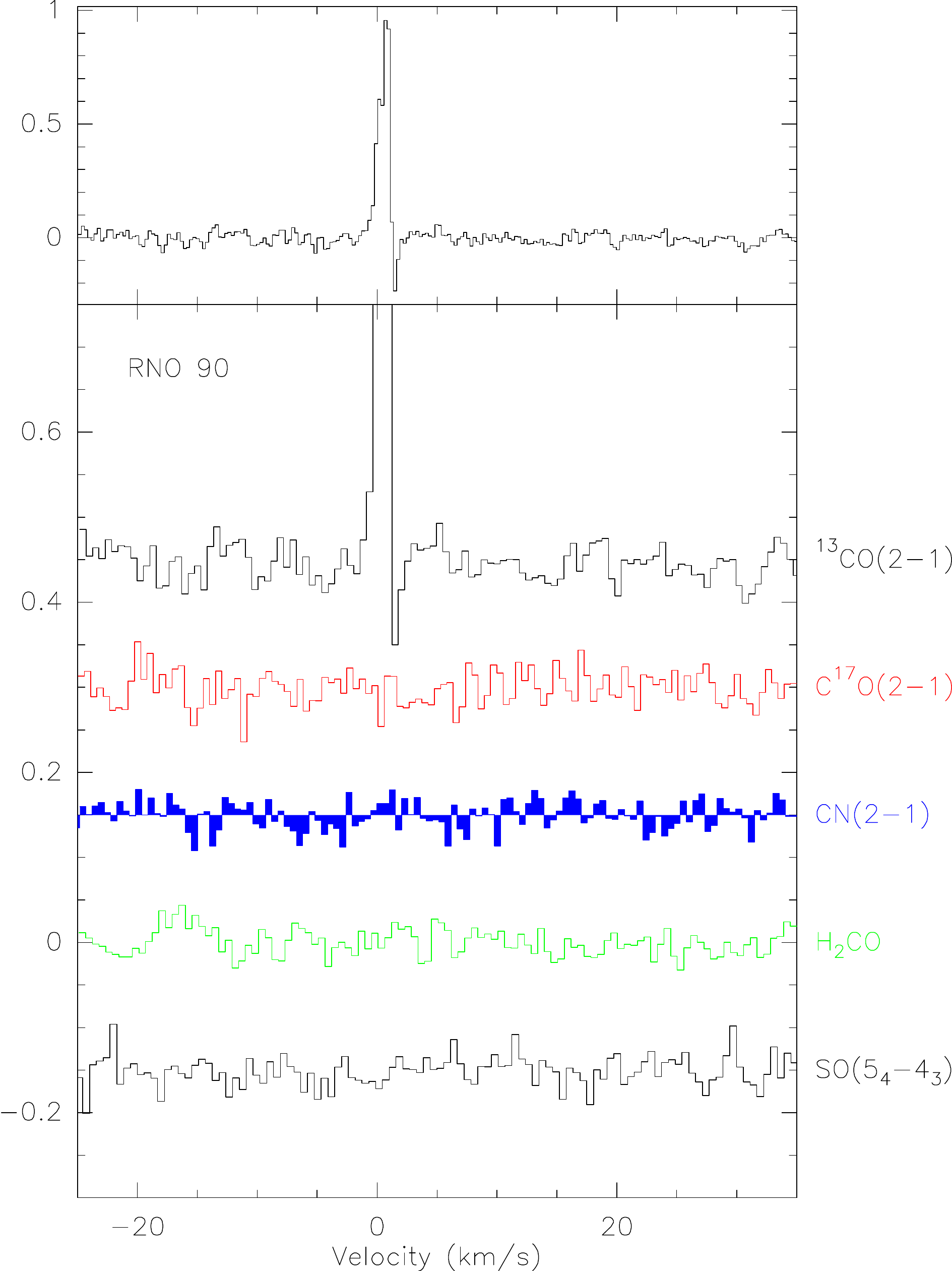}
\caption{Lines toward RNO 90.}
\label{fig:spe-RNO_90}
\end{figure}
\begin{figure}
\includegraphics[width=8.0cm]{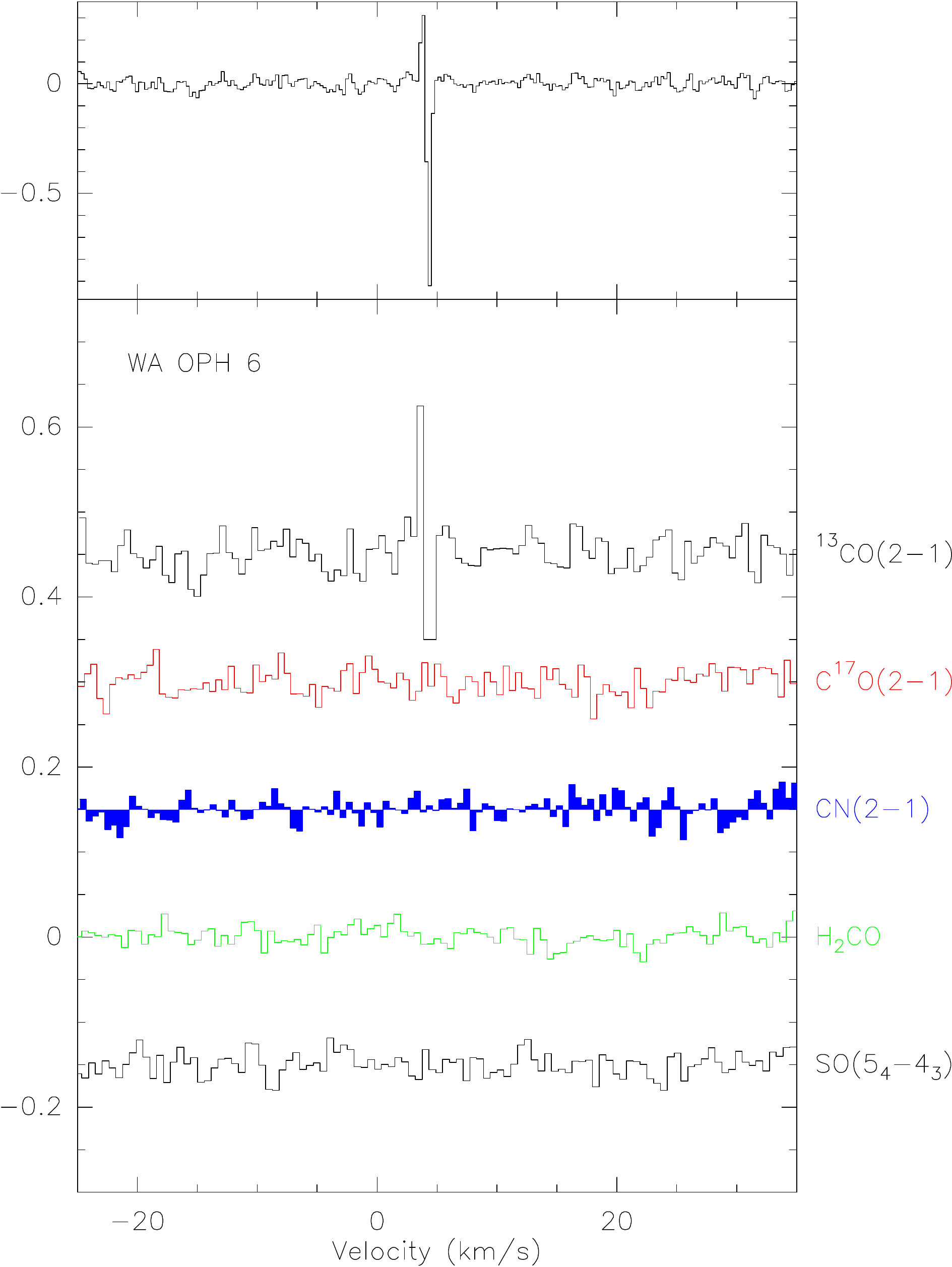}
\caption{Lines toward Wa Oph 6.}
\label{fig:spe-WA_OPH_6}
\end{figure}
\clearpage
\begin{figure}
\includegraphics[width=8.0cm]{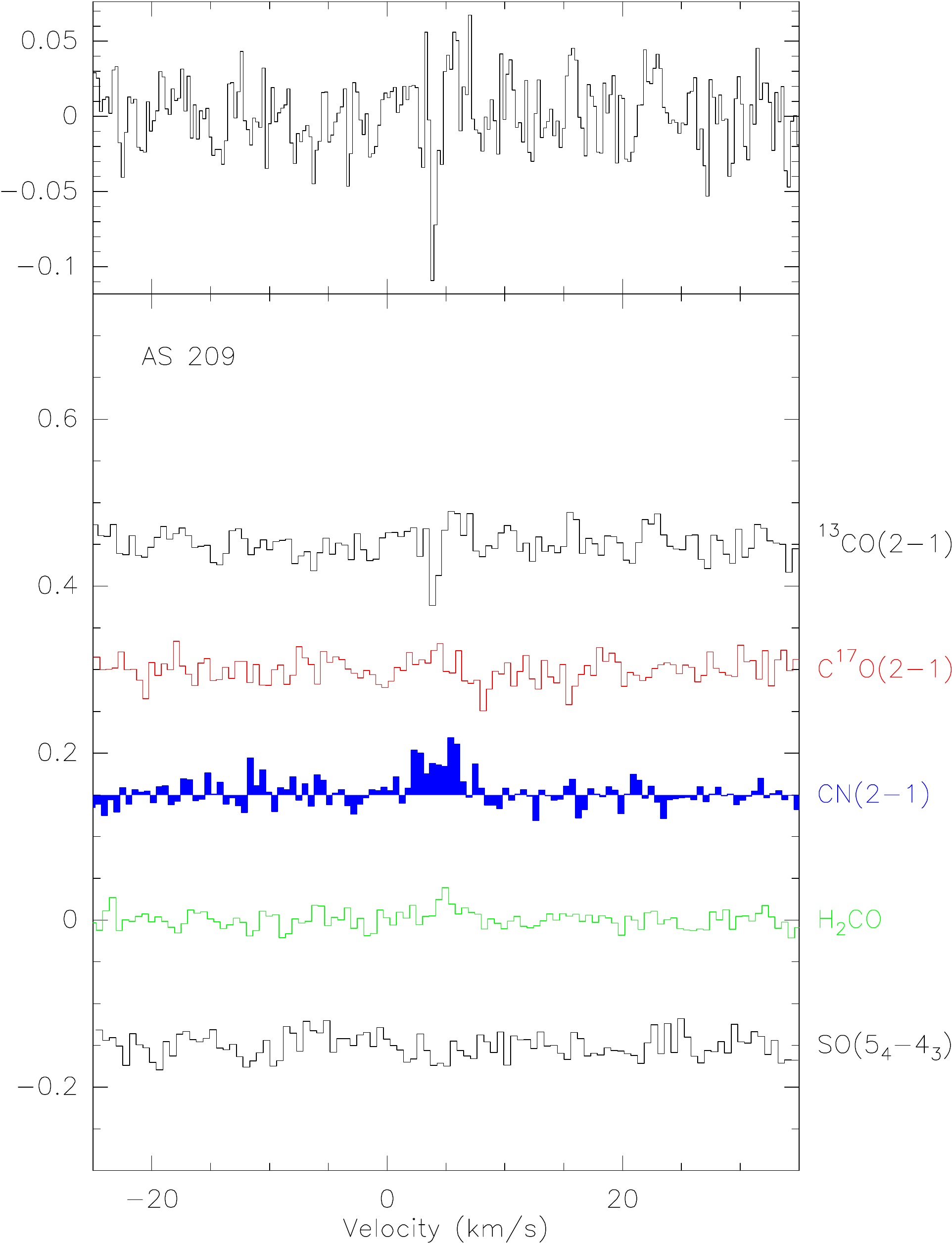}
\caption{Lines toward AS 209.}
\label{fig:spe-AS_209}
\end{figure}
\begin{figure}
\includegraphics[width=8.0cm]{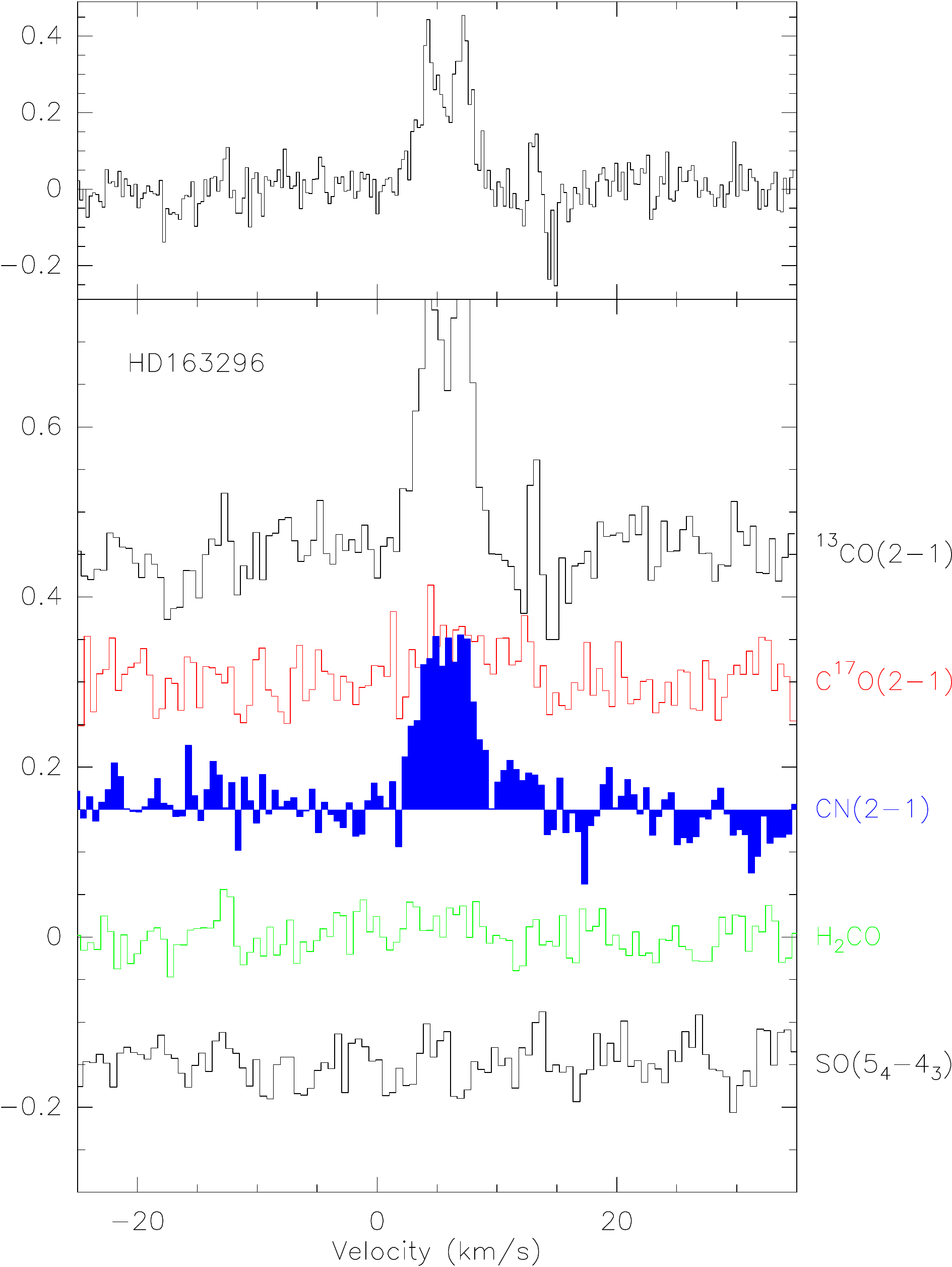}
\caption{Lines toward HD 163296.}
\label{fig:spe-HD163296}
\end{figure}

\end{document}